\documentclass[12pt,english]{article}
\usepackage{tgtermes}
\usepackage{helvet}
\usepackage[lf]{FiraMono}

\usepackage[T1]{fontenc}
\usepackage[latin9]{inputenc}
\usepackage[a4paper]{geometry}
\usepackage{array}
\usepackage{mathtools}
\usepackage{multirow}
\usepackage{amsmath}
\usepackage{amssymb}
\usepackage{graphicx}
\usepackage[authoryear]{natbib}

\makeatletter

\providecommand{\tabularnewline}{\\}

\usepackage{pdfpages}
\usepackage{babel}

\makeatother

\usepackage{babel}
\begin{document}
\title{Quasi-Bayesian Local Projection Instrumental-Variables Method: Application
to Renewable Energy and Electricity Prices}
\author{Masahiro Tanaka\thanks{Corresponding author. Faculty of Economics, Fukuoka University, Fukuoka,
Japan. Address: 8-19-1, Nanakuma, Jonan, Fukuoka, Japan 814-0180.
E-mail: m.tanaka.tt@fukuoka-u.ac.jp.}}
\maketitle
\begin{abstract}
This paper introduces a quasi-Bayesian approach for local projection
instrumental-variables (LP--IV) estimation. It builds a moment-based
quasi-posterior using the generalized method of moments (GMM) objective
and applies a roughness-penalty prior to smooth impulse responses
over different horizons. The approach maintains the key first-order
features of traditional LP--IV methods, while enhancing stability
in finite samples and allowing for joint inference through simultaneous
bands. Simulations indicate that this regularization decreases root
mean squared error compared to standard GMM, especially at medium
and longer horizons. An application to Danish electricity markets
highlights the method's practical usefulness.\bigskip{}
\end{abstract}



\section{Introduction}

Estimating dynamic causal effects is essential in applied econometrics.
Local projections assess impulse responses by employing regressions
specific to each horizon \citep{Jorda2005}.\footnote{See \citet{Jorda2023,Jorda2025,Inoue2026} for a review.}
When the treatment variable is endogenous, dynamic causal effects
can be identified by combining local projections with instrumental
variables, which results in LP--IV estimators \citep{Jorda2015,Ramey2016,Stock2018,Ramey2018}.
Since LP--IV does not require defining a complete dynamic system,
it provides a flexible framework for causal impulse-response analysis
across macroeconomics, finance, and applied microeconomics. A key
limitation of LP--IV is that estimates based on finite samples can
vary greatly across different horizons. Since the impulse response
is estimated independently at each horizon, sampling noise can create
irregular response patterns, particularly at medium- and long-term
horizons. This issue is especially pronounced in high-frequency contexts
with volatile outcomes and low signal-to-noise ratios. In these cases,
unrestricted LP--IV estimates might be hard to interpret, even when
economic adjustment is presumed to be smooth. This paper introduces
a quasi-Bayesian LP--IV approach that regularizes impulse responses
over different horizons. It considers the impulse response as a function
of the horizon and applies a roughness-penalty prior to reduce differences
between neighboring horizons. The quasi-posterior is derived from
the generalized method of moments (GMM) objective function based on
the IV moment conditions, following the moment-based quasi-Bayesian
framework of \citet{Chernozhukov2003}. The estimator integrates IV
identification with structured regularization, without requiring a
full likelihood specification.

The proposed approach has two main features. First, it offers a quasi-Bayesian
formulation of LP--IV. Unlike existing Bayesian and quasi-Bayesian
methods for local projections, such as \citet{Tanaka2020} and \citet{Ferreira2025},
which are designed for setups where impulse responses are identified
through narrative restrictions or vector autoregressions \citep{Ramey2016},
these frameworks do not directly include IV moment restrictions. By
deriving the quasi-posterior from the GMM objective function, this
method integrates LP--IV moment conditions directly into Bayesian-like
inference. Second, assuming certain regularity conditions, the quasi-posterior
mean is asymptotically equivalent to the standard GMM estimator. Additionally,
the roughness-penalty prior improves finite-sample estimates by sharing
information across horizons. This prior is a proper Gaussian Markov
random field with a hierarchical scale parameter, enabling data-driven
regularization. Unlike frequentist smoothing techniques for LPs \citep{Barnichon2018,Barnichon2019,El-Shagi2019},
this approach jointly estimates the smoothing and model parameters.

Joint inference plays a crucial role in impulse-response analysis.
An impulse response represents a trajectory showing how an outcome
changes over time after an intervention, rather than just a set of
isolated scalar parameters. Therefore, uncertainty about this trajectory
requires simultaneous inference across all points, not just individual,
pointwise estimates. As highlighted by \citet{Jorda2023,Jorda2025},
relying solely on pointwise intervals can be misleading because response
estimates are correlated across different horizons. Consequently,
inference about the overall shape of the response should incorporate
cross-horizon dependence.

The proposed quasi-Bayesian framework also facilitates joint inference
by estimating the entire sequence of LP--IV coefficients as a combined
parameter vector. In our empirical application, we present pointwise
confidence intervals and simultaneous confidence bands derived from
a sandwich covariance matrix for the stacked moment conditions. Therefore,
the baseline uncertainty measures are frequentist, while the quasi-Bayesian
approach helps stabilize the estimation of the impulse-response path.
This approach differs from \citet{Ferreira2025}, which assesses uncertainty
on an equation-by-equation basis. Unlike such individual equation
methods, the stacked moment-based formulation used here simplifies
simultaneous inference for LP--IV response paths.

A simulation study indicates that the quasi-Bayesian estimator using
a flat prior closely matches the classical GMM estimator, reinforcing
the connection between the quasi-posterior and traditional LP--IV
estimation. When employing the roughness-penalty prior, the estimator
applies moderate shrinkage and generally lowers RMSE in the baseline
and block-diagonal weighting configurations, particularly at medium
and longer horizons.

The application analyzes how renewable generation affects Danish wholesale
electricity prices over time. Using daily data from Western Denmark
(DK1) and Eastern Denmark (DK2), it models wind and solar generation
based on weather-driven renewable potential derived from ERA5 reanalysis
data. Findings reveal that wind generation significantly lowers prices
in the short term in both zones, aligning with the merit-order effect
described in electricity market studies. The effect of wind is more
pronounced in DK1 and diminishes over the following days. Solar effects
are smaller and less precisely measured in daily data. Numerous robustness
tests, including alternative LP models, lag lengths, seasonal controls,
prior hyperparameters, lagged renewable states, and placebo exercises,
consistently support these core findings.

The paper makes contributions to two key areas. First, it advances
the econometric field of impulse-response estimation by introducing
a moment-based quasi-Bayesian approach for regularized LP--IV estimation.
Unlike methods that rely on post-estimation smoothing or impose parametric
restrictions on the impulse response, this new approach integrates
smoothness directly into the estimation process while maintaining
IV-based identification. Second, it adds to the empirical research
on renewable energy and electricity prices by analyzing how weather-driven
fluctuations in renewable supply influence dynamic price responses.
The findings indicate that renewable integration affects both immediate
prices and short-term price movements, which has important implications
for balancing markets, storage incentives, and risk management.

The remainder of the paper is structured as follows. Section 2 discusses
the economic background and empirical motivation. Section 3 outlines
the quasi-Bayesian LP--IV framework. Section 4 details the simulation
study. Section 5 covers the empirical application to renewable generation
and electricity prices. Finally, Section 6 offers concluding remarks.

\section{Economic Background and Empirical Motivation}

Electricity markets are ideal for analyzing how supply shocks change
over time. The rise of variable renewable energy sources, like wind
and solar, has caused significant weather-related fluctuations in
supply, affecting price volatility, investment in generation, and
system reliability. For policymakers, understanding the effect of
renewable generation on prices is key for shaping energy and climate
strategies, such as renewable subsidies, carbon pricing, and capacity
mechanisms. Market participants also benefit from knowing how renewable
supply shocks spread over time, aiding in operational planning, risk
management, and investment decisions. Due to high-frequency data and
clearly defined market structures, electricity markets are particularly
suited for studying how external supply changes lead to dynamic price
responses.

Extensive research shows that renewable energy reduces wholesale electricity
prices via the merit-order effect \citep{Sensfuss2008,Wuerzburg2013,Clo2015}.
Wind and solar power have low marginal costs and can displace more
expensive thermal generation, thereby lowering prices. While this
immediate effect is well known, less is understood about how renewable
supply shocks influence prices over later periods. This dynamic aspect
is important for shaping storage incentives, reserve requirements,
and cross-border trade. Persistence might stem from thermal units'
ramping and startup constraints \citep{Borenstein2002}, hydroelectric
and storage options enable intertemporal substitution; expectations
about short-term supply influence bidding strategies; and interconnections
transmit shocks across regions and trading periods \citep{Green2011}.
More generally, renewable intermittency poses dynamic adjustment challenges
beyond simple supply-demand interactions \citep{Joskow2011}.

From a policy standpoint, the focus is not just on the immediate price
effect of renewable generation but also on the entire adjustment process
after a supply shock. Let $\gamma_{\left(h\right)}$ denote the causal
effect of a one-unit increase in renewable generation at time $t$
on the electricity price $h$ periods ahead. The sequence $\left\{ \gamma_{\left(h\right)}\right\} _{h=0}^{H}$
characterizes how prices respond over time and helps distinguish transitory
fluctuations from more persistent effects that may influence investment
incentives and system reliability. This response path is also important
for market participants in their price forecasting, risk management,
as well as production and trading decisions.

A significant empirical challenge is that observed renewable generation
might be correlated with other factors influencing electricity prices,
such as demand patterns, seasonal fluctuations, and broader weather
conditions. To accurately evaluate policies, it is essential to isolate
exogenous variations in renewable supply. Recent studies have tackled
this by employing meteorological variables as instruments for renewable
generation. Wind speed and solar irradiance directly affect renewable
output and, when controlling for demand, weather, and seasonal factors,
offer plausibly exogenous supply variation \citep{Staffell2016,Pfenninger2016}.
Research using weather-based instruments demonstrates economically
significant effects of renewable generation on electricity prices
and related market outcomes \citep{Hirth2013}.

Empirical analysis in this context faces several statistical challenges.
First, renewable supply shocks have effects that change over time,
necessitating estimates at multiple horizons. Second, electricity
prices show significant high-frequency volatility, which can lead
to imprecise horizon-specific estimates \citep{Weron2014}. Third,
although prices may spike sharply initially, economic adjustments
following the shock are generally expected to develop smoothly across
nearby horizons. These characteristics suggest a method that captures
the entire dynamic response while smoothing out sampling noise, aligning
with realistic economic adjustment patterns.

\section{Method}

\subsection{LP--IV}

The objective is to estimate the dynamic causal effect of a treatment
$r_{t}$ on an outcome $y_{t}$. In the empirical application, $r_{t}$
denotes renewable generation and $y_{t}$ denotes the wholesale electricity
price. The parameter of interest is the impulse-response sequence
$\left\{ \gamma_{\left(h\right)}\right\} _{h=0}^{H}$ , where $\gamma_{\left(h\right)}$
measures the effect of a treatment shock at time $t$ on the outcome
$h$ periods ahead. Estimation encounters two main challenges: the
treatment may be endogenous, and horizon-specific estimates of $\gamma_{\left(h\right)}$
can be noisy, especially at longer horizons. This section presents
a framework that tackles these challenges by integrating IV identification
with regularized quasi-Bayesian inference.

We determine the impulse response by conducting a series of horizon-specific
regressions. For clarity, this section discusses a single treatment
variable $r_{t}$. However, the framework can handle multiple treatment
variables, as demonstrated in the empirical application. Throughout
the paper, we focus on just-identified models where the number of
instruments matches the number of regressors. This restriction simplifies
both estimation and computation and enables the initial IV estimator
to be expressed in a closed form. We consider two LP--IV specifications.
The first is the level specification:

\[
y_{t+h}=\gamma_{\left(h\right)}r_{t}+\eta_{\left(h\right)}+\sum_{l=1}^{L}\mu_{\left(h\right),l}y_{t-l}+\boldsymbol{\delta}_{\left(h\right)}{}^{\top}\tilde{\boldsymbol{x}}_{t}+e_{\left(h\right),t},
\]
where $\tilde{\boldsymbol{x}}_{t}$ is a vector of control variables
and $e_{\left(h\right),t}$ is an error term. The coefficient $\gamma_{\left(h\right)}$
is the target causal effect at horizon $h$, while $\eta_{\left(h\right)}$,
$\left\{ \mu_{\left(h\right),l}\right\} _{l=1}^{L}$, and $\boldsymbol{\delta}_{\left(h\right)}$
are nuisance parameters.

The second is the long-differenced (LD) specification, which replaces
the dependent variable $y_{t+h}$ with $y_{t+h}-y_{t-1}$ and uses
lagged first differences of the outcome as controls \citep{Stock2018}:
\[
y_{t+h}-y_{t-1}=\gamma_{\left(h\right)}r_{t}+\eta_{\left(h\right)}+\sum_{l=1}^{L}\mu_{\left(h\right),l}\Delta y_{t-l}+\boldsymbol{\delta}_{\left(h\right)}{}^{\top}\tilde{\boldsymbol{x}}_{t}+e_{\left(h\right),t},
\]
where $\Delta y_{t-l}=y_{t-l}-y_{t-l-1}$. Under no anticipation,
$y_{t-1}$ is unaffected by the treatment shock at $t$, so the LD
coefficient $\gamma_{\left(h\right)}$ can be interpreted as the impulse
response of $y_{t+h}$ relative to the pre-shock baseline. As the
LD specification can reduce bias and improve coverage in finite samples
\citep{Piger2025}, we use the LD specification as the baseline in
the analysis below.

We define the $J$-dimensional vectors of regressors and corresponding
coefficients as 
\[
\boldsymbol{x}_{t}=\left(r_{t},1,\Delta y_{t-1},\cdots,\Delta y_{t-L},\tilde{\boldsymbol{x}}_{t}^{\top}\right)^{\top},
\]
and 
\[
\boldsymbol{\theta}_{\left(h\right)}=\left(\gamma_{\left(h\right)},\eta_{\left(h\right)},\mu_{\left(h\right),1},...,\mu_{\left(h\right),L},\boldsymbol{\delta}_{\left(h\right)}{}^{\top}\right)^{\top}.
\]
The horizon-$h$ regression can then be written compactly as 
\[
y_{t+h}-y_{t-1}=\boldsymbol{\theta}_{\left(h\right)}^{\top}\boldsymbol{x}_{t}+e_{\left(h\right),t}.
\]

Inference is based on the IV moment conditions. For each horizon,
the structural error is assumed to be orthogonal to a $J$-dimensional
instrument vector $\boldsymbol{z}_{t}$. Stacking the horizon-specific
coefficient vectors gives 
\[
\boldsymbol{\theta}=\left(\boldsymbol{\theta}_{\left(0\right)}^{\top},...,\boldsymbol{\theta}_{\left(H\right)}^{\top}\right)^{\top}.
\]
We define the stacked moment function: 
\[
\boldsymbol{m}_{t}\left(\boldsymbol{\theta}\right)=\left(\boldsymbol{m}_{\left(0\right),t}\left(\boldsymbol{\theta}\right)^{\top},...,\boldsymbol{m}_{\left(H\right),t}\left(\boldsymbol{\theta}\right)^{\top}\right)^{\top}.
\]
Here, each $\boldsymbol{m}_{\left(h\right),t}\left(\boldsymbol{\theta}\right)$
is the horizon-$h$ IV moment contribution. Specifically, we define
\[
\boldsymbol{m}_{\left(h\right),t}\left(\boldsymbol{\theta}\right)=\left(y_{t+h}-y_{t-1}-\boldsymbol{\theta}_{\left(h\right)}^{\top}\boldsymbol{x}_{t}\right)\boldsymbol{z}_{t}.
\]
Because each horizon-specific regression generates $J$ orthogonality
conditions, the stacked moment vector has dimension $K=J\left(H+1\right)$.
Therefore, the population moment condition is 
\[
\mathbb{E}\left[\boldsymbol{m}_{t}\left(\boldsymbol{\theta}_{0}\right)\right]=\boldsymbol{0}_{K},
\]
where $\boldsymbol{0}_{K}$ denotes the $K$-dimensional vector of
zeros and $\boldsymbol{\theta}_{0}$ is the true parameter vector.

\subsection{Moment-based quasi-Bayesian inference}

Rather than specifying a full likelihood for the observed variables,
we use a GMM-based quasi-posterior to obtain a regularized moment-based
estimator. Following \citet{Chernozhukov2003}, we define 
\[
\pi\left(\boldsymbol{\theta}\right)=\frac{\exp\left\{ -\frac{T}{2}\bar{\boldsymbol{m}}\left(\boldsymbol{\theta}\right)^{\top}\boldsymbol{W}\bar{\boldsymbol{m}}\left(\boldsymbol{\theta}\right)\right\} p\left(\boldsymbol{\theta}\right)}{\int\exp\left\{ -\frac{T}{2}\bar{\boldsymbol{m}}\left(\boldsymbol{\theta}\right)^{\top}\boldsymbol{W}\bar{\boldsymbol{m}}\left(\boldsymbol{\theta}\right)\right\} p\left(\boldsymbol{\theta}\right)d\boldsymbol{\theta}},
\]
where 
\[
\bar{\boldsymbol{m}}\left(\boldsymbol{\theta}\right)=\frac{1}{T}\sum_{t=1}^{T}\boldsymbol{m}_{t}\left(\boldsymbol{\theta}\right),
\]
$\boldsymbol{W}$ is a positive definite weighting matrix, and $p\left(\boldsymbol{\theta}\right)$
is a prior. Throughout, $T$ denotes the number of usable projection-origin
observations after trimming initial lags and terminal leads. The quasi-posterior
uses the moment restrictions directly and does not require a fully
specified likelihood for data generation.

It is helpful to differentiate between the flat-prior and regularized
cases. Under the flat prior $p\left(\boldsymbol{\theta}\right)\propto1$,
the quasi-posterior is centered at the minimizer of the quadratic
GMM criterion. In the linear exactly identified LP--IV model considered
here, this estimator coincides with the conventional IV/GMM estimator.
The flat-prior quasi-posterior, therefore, provides a direct moment-based
representation of classical GMM inference. Under the roughness-penalty
prior introduced in Section 3.3, the quasi-posterior mean is generally
not equal to the conventional GMM estimator in finite samples. The
prior regularizes the stacked coefficient vector by shrinking horizon-indexed
coefficient paths toward smoother trajectories. Hence the resulting
estimator trades off fit to the IV moment conditions against smoothness
across horizons. This finite-sample regularization serves as the mechanism
enabling the proposed method to reduce sampling variability compared
to unrestricted LP--IV.

The quasi-likelihood is formulated from the quadratic GMM objective
function linked to IV moment conditions \citep{Hansen1982}. The resulting
quasi-posterior merges insights from these moments with prior knowledge
of the parameter vector. Since it relies on moment restrictions rather
than a comprehensive likelihood, this approach does not necessitate
specifying the data's joint distribution and avoids issues stemming
from likelihood misspecification. Alternatively, the quasi-posterior
can be interpreted as a Gibbs or generalized Bayesian posterior derived
from the quadratic GMM loss \citep{Jiang2008,Bissiri2016}, connecting
it to Bayesian inference based on estimating equations and moment
restrictions \citep{Kim2002,Yin2009,Liao2011,Li2016}.

We assume that the marginal prior satisfies the standard local prior-negligibility
condition
\[
\log p\left(\boldsymbol{\theta}_{0}+\frac{c}{\sqrt{T}}\right)-\log p\left(\boldsymbol{\theta}_{0}\right)=o\left(1\right)
\]
uniformly over compact sets of $c$. Therefore, the quasi-posterior
mean $\hat{\boldsymbol{\theta}}$ is first-order equivalent to the
conventional GMM estimator \citep{Chernozhukov2003,Hong2021}: 
\[
\sqrt{T}\left(\hat{\boldsymbol{\theta}}-\boldsymbol{\theta}_{0}\right)\rightarrow\mathcal{N}\left(\boldsymbol{0},\boldsymbol{V}\right),
\]
\[
\boldsymbol{V}=\left(\boldsymbol{G}^{\top}\boldsymbol{W}\boldsymbol{G}\right)^{-1}\boldsymbol{G}^{\top}\boldsymbol{W}\boldsymbol{\Sigma}_{0}\boldsymbol{W}\boldsymbol{G}\left(\boldsymbol{G}^{\top}\boldsymbol{W}\boldsymbol{G}\right)^{-1}.
\]
Here $\boldsymbol{G}=\mathbb{E}\left[\partial\boldsymbol{m}_{t}\left(\boldsymbol{\theta}_{0}\right)/\partial\boldsymbol{\theta}^{\top}\right]$
is the Jacobian of the population moment condition $\boldsymbol{m}_{t}\left(\boldsymbol{\theta}_{0}\right)$,
and $\boldsymbol{\Sigma}_{0}$ is the covariance matrix of $\boldsymbol{m}_{t}\left(\boldsymbol{\theta}_{0}\right)$.
We use the contemporaneous covariance $\boldsymbol{\Sigma}_{0}=\mathrm{Var}\left[\boldsymbol{m}_{t}\left(\boldsymbol{\theta}_{0}\right)\right]$,
rather than a HAR long-run covariance matrix, as the baseline. The
motivation is close to \citet{MontielOlea2021}, who show that lag-augmented
local projections do not require correction for serial correlation
in the multi-step residuals: although the projection residuals are
serially correlated, the relevant regression scores are serially uncorrelated
under their assumptions, so heteroskedasticity-robust standard errors
suffice. In the present LP--IV formulation, the analogous objects
are the IV moment scores $\boldsymbol{m}_{\left(h\right),t}\left(\boldsymbol{\theta}\right)=\boldsymbol{z}_{t}e_{\left(h\right),t}\left(\boldsymbol{\theta}\right)$.
Accordingly, the baseline asymptotic variance uses the covariance
of these stacked moment scores rather than a HAR long-run covariance
estimator. This choice is an identifying and inferential convention
for the baseline specification; it is distinct from the cross-horizon
covariance captured by stacking $\boldsymbol{m}_{t}\left(\boldsymbol{\theta}\right)$.

In the LP--IV model, $\boldsymbol{G}$ is estimated by its sample
analogue, 
\[
\hat{\boldsymbol{G}}=\boldsymbol{I}_{H+1}\otimes\left(-\frac{1}{T}\boldsymbol{Z}^{\top}\boldsymbol{X}\right),
\]
where 
\[
\boldsymbol{X}=\left(\boldsymbol{x}_{1},...,\boldsymbol{x}_{T}\right)^{\top},\quad\boldsymbol{Z}=\left(\boldsymbol{z}_{1},...,\boldsymbol{z}_{T}\right)^{\top},
\]
$\boldsymbol{I}_{H+1}$ is the $\left(H+1\right)\times\left(H+1\right)$
identity matrix, and $\otimes$ denotes the Kronecker product. This
first-order equivalence does not imply equality in finite samples:
the quasi-posterior mean under an informative prior may differ from
the GMM estimator.

The quasi-Bayesian approach offers two main benefits. First, it provides
a straightforward way to include structured regularization through
hierarchical priors. Second, since it estimates the entire sequence
of horizon-specific coefficients together, it naturally supports joint
uncertainty quantification for the full IRF. In the basic implementation
described below, uncertainty is calibrated using sandwich-based frequentist
bands centered on the quasi-posterior mean, while the hierarchical
prior helps regularize the point estimate by determining the level
of cross-horizon smoothing based on the data.

For computational stability, we fix the weighting matrix $\boldsymbol{W}$
throughout the simulation. This follows \citet{Chernozhukov2003}
and avoids recomputing the covariance matrix of the moment conditions
at each draw. In the reported inference, uncertainty is calibrated
using the asymptotic sandwich covariance described below. We set $\boldsymbol{W}$
equal to the inverse of the empirical covariance matrix of the moment
functions evaluated at an initial consistent estimator. In the exactly
identified specification considered here, this initial estimator is
\[
\boldsymbol{\theta}^{*}=\textrm{vec}\left(\left(\boldsymbol{Z}^{\top}\boldsymbol{X}\right)^{-1}\boldsymbol{Z}^{\top}\boldsymbol{Y}\right),
\]
where 
\[
\boldsymbol{Y}=\left(\begin{array}{ccc}
y_{1}-y_{0} & \cdots & y_{1+H}-y_{0}\\
\vdots & \cdot & \vdots\\
y_{T}-y_{T-1} & \cdots & y_{T+H}-y_{T-1}
\end{array}\right),
\]
and $\textrm{vec}\left(\cdot\right)$ denotes the column-wise vectorization.

When $K$ is large, estimating and inverting the entire covariance
matrix of the moment functions can become numerically unstable. In
these situations, we opt for a block-diagonal weighting matrix that
spans different horizons, 
\[
\boldsymbol{W}=\textrm{blkdiag}\left(\boldsymbol{W}_{\left(0\right)},...,\boldsymbol{W}_{\left(H\right)}\right),
\]
where $\boldsymbol{W}_{\left(h\right)}$ is the inverse of the empirical
covariance matrix of $\boldsymbol{m}_{\left(h\right),t}\left(\boldsymbol{\theta}\right)$.
This approximation maintains the within-horizon weighting to prevent
unstable inversion of the entire stacked covariance matrix. The block-diagonal
restriction applies solely to the weighting matrix for estimation.
Meanwhile, the sandwich covariance matrix used for inference is derived
from the complete stacked moment vector, preserving the cross-horizon
dependence.

\subsection{Proper roughness-penalty prior}

In impulse-response analysis, the coefficients across different horizons
form a dynamic adjustment path rather than just a set of unrelated
scalar parameters. In many cases, nearby horizons are expected to
contain similar information because economic adjustments are driven
by persistent mechanisms such as production constraints, storage,
expectations, or institutional frictions. Therefore, smoothness across
horizons can serve as a scientifically justified regularization technique:
it helps reduce sampling noise while still allowing the data to shape
the response's level, slope, and form.

For each covariate $j$, define the horizon-indexed coefficient path
\[
\boldsymbol{\theta}_{j}=\left(\theta_{\left(0\right),j},...,\theta_{\left(H\right),j}\right)^{\top}.
\]
We assign the conditional Gaussian prior 
\[
\boldsymbol{\theta}_{j}|\tau_{j}\sim\mathcal{N}\left(\boldsymbol{0}_{H+1},\;\tau_{j}^{2}\boldsymbol{Q}^{-1}\right),
\]
where $\tau_{j}>0$ controls the overall scale of the path. The matrix
$\boldsymbol{Q}$ is a baseline smoothing precision matrix over the
projection horizon and is specified as
\[
\boldsymbol{Q}=\boldsymbol{D}^{\top}\boldsymbol{D}+\frac{8}{\rho^{2}}\boldsymbol{I}_{H+1},
\]
where $\boldsymbol{D}$ is the first-order difference matrix of dimension
$H\times\left(H+1\right)$ and $\rho>0$ is a correlation-range parameter.
The prior penalizes both adjacent differences and the overall level
of the coefficient path. The term $\boldsymbol{D}^{\top}\boldsymbol{D}$
regularizes the shape of the coefficient path by penalizing adjacent
differences, while leaving constant paths unpenalized. The ridge term
makes the prior proper by penalizing this constant component that
would otherwise remain unregularized, thus directly shrinking the
overall path level toward zero. In the implementation, the prior is
applied to each horizon-indexed coefficient path, including both nuisance
coefficients and the IRF coefficients, although the latter are the
main objects of interest.

This prior can be viewed as a proper Gaussian Markov random field
over the projection horizon \citep{Lindgren2011}. It is also closely
connected to the discrete approximation of a Matérn-type Gaussian
field. In the notation of a continuous horizon, the related field
corresponds to the stochastic differential operator
\[
\left(\frac{8}{\rho^{2}}-\frac{\partial^{2}}{\partial h^{2}}\right)f\left(h\right)=\mathcal{W}\left(h\right),
\]
where $\mathcal{W}\left(h\right)$ denotes Gaussian white noise. The
precision matrix is a finite-difference analog of the continuous precision
operator $8/\rho^{2}-\partial^{2}/\partial h^{2}$, up to boundary
and grid-spacing constants. The correlation-range parameter $\rho$
has a quantitative interpretation. For an interior point of a long
horizon grid, the prior correlation between two coefficients separated
by $\left|h-h^{\prime}\right|$ horizons is approximately 
\[
\mathrm{Corr}\left(\theta_{\left(h\right),j},\theta_{\left(h^{\prime}\right),j}\right)\approx\exp\left(-\frac{\sqrt{8}}{\rho}\left|h-h^{\prime}\right|\right).
\]
Thus, larger values of $\rho$ imply slower decay of prior dependence
and smoother coefficient paths. Quantitatively, the prior correlation
falls to approximately 0.5 after $0.25\rho$ horizons, to $1/e\approx0.37$
after $0.35\rho$ horizons, and to 0.1 after $0.8\rho$ horizons.

The limiting case $\rho\rightarrow\infty$ corresponds to the intrinsic
first-order random-walk prior \citep{Lang2004}, which penalizes roughness;
however, it is improper. For any finite $\rho$, the ridge component
makes $\boldsymbol{Q}$ positive definite and defines a valid prior
distribution for the full coefficient path.

To enable data-driven assessment of shrinkage strength, we assign
a hierarchical prior to the scale parameter, 
\[
\tau_{j}\sim\mathcal{C}^{+}\left(0,\kappa\right),
\]
where $\mathcal{C}^{+}\left(0,\kappa\right)$ denotes a half-Cauchy
distribution with scale parameter $\kappa$. This heavy-tailed prior
enables large coefficient paths when supported by data, while pushing
weakly supported paths toward zero and promoting smoother variation
across horizons \citep{Polson2012}.

The hierarchical prior regularizes the IRF across different horizons,
with the smoothing degree estimated together with the coefficients.
It offers a systematic method for the regularized estimation of horizon-indexed
parameters identified through moments. The prior functions solely
as a regularization tool, while the dynamic causal effects are identified
by the IV moment conditions.

A related class of priors was introduced by \citet{Tanaka2020}. That
prior is improper, corresponding to the limiting case $\rho\rightarrow\infty$,
whereas the present formulation is proper for any finite $\rho$.
Unlike frequentist smoothing approaches to LPs \citep{Barnichon2018,Barnichon2019,El-Shagi2019},
the quasi-Bayesian framework treats the prior scale parameters as
unknown, rather than selecting the amount of smoothing outside the
joint estimation problem.

\subsection{Computation}

We simulate from the quasi-posterior using a Gibbs sampler. The quasi-posterior
kernel of $\boldsymbol{\theta}$ conditional on $\boldsymbol{\tau}$
can be written as 
\[
\pi\left(\boldsymbol{\theta}|\boldsymbol{\tau}\right)\propto\exp\left\{ -\frac{T}{2}\left(\boldsymbol{\theta}-\boldsymbol{\theta}^{*}\right)^{\top}\hat{\boldsymbol{G}}^{\top}\boldsymbol{W}\hat{\boldsymbol{G}}\left(\boldsymbol{\theta}-\boldsymbol{\theta}^{*}\right)\right\} p\left(\boldsymbol{\theta}|\boldsymbol{\tau}\right).
\]
Combining the quadratic GMM loss and conditional Gaussian prior yields
\[
\boldsymbol{\theta}|\boldsymbol{\tau}\sim\mathcal{N}\left(\boldsymbol{\Omega}\boldsymbol{\Upsilon}\boldsymbol{\theta}^{*},\;\boldsymbol{\Omega}\right),
\]
where 
\[
\boldsymbol{\Omega}=\left(\boldsymbol{\Upsilon}+\boldsymbol{\Pi}\right)^{-1},\quad\boldsymbol{\Upsilon}=T\hat{\boldsymbol{G}}^{\top}\boldsymbol{W}\hat{\boldsymbol{G}},\quad\boldsymbol{\Pi}=\boldsymbol{Q}\otimes\textrm{diag}\left(\tau_{1}^{-2},...,\tau_{J}^{-2}\right).
\]
Sampling from this Gaussian distribution leverages the sparse structure
of the prior precision matrix $\boldsymbol{Q}$. We implement the
Gaussian Markov random field simulation algorithm of \citet{Rue2001},
which improves computational efficiency.

The scale parameters are updated using the half-Cauchy representation
as a scale mixture of inverse-gamma distributions \citep{Wand2011}:
\[
\tau_{j}^{2}|\nu_{j}\sim\mathcal{IG}\left(\frac{1}{2},\frac{1}{\nu_{j}}\right),\quad\nu_{j}\sim\mathcal{IG}\left(\frac{1}{2},\frac{1}{\kappa^{2}}\right),
\]
where $\nu_{j}$ is an auxiliary random variable and $\mathcal{IG}\left(a,b\right)$
denotes an inverse gamma distribution with shape parameter $a$ and
rate parameter $b$. The corresponding full conditional distributions
are derived as in \citet{Makalic2015}: 
\begin{eqnarray*}
\tau_{j}^{2}|\boldsymbol{\theta}_{j},\nu_{j} & \sim & \mathcal{IG}\left(\frac{H+2}{2},\;\frac{1}{\nu_{j}}+\frac{1}{2}\boldsymbol{\theta}_{j}^{\top}\boldsymbol{Q}\boldsymbol{\theta}_{j}\right),\\
\nu_{j}|\tau_{j}^{2} & \sim & \mathcal{IG}\left(1,\;\frac{1}{\kappa^{2}}+\frac{1}{\tau_{j}^{2}}\right).
\end{eqnarray*}

The quasi-posterior mean is estimated by averaging the simulated draws
of $\boldsymbol{\theta}$. While the MCMC algorithm samples both the
coefficient vector and the smoothing-scale parameters, our primary
uncertainty summaries are based on a frequentist approach rather than
a Bayesian posterior. Specifically, we present sandwich standard errors
calculated around the quasi-posterior mean. This approach distinguishes
the regularization employed for point estimation from the calibration
of sampling uncertainty. Consequently, the derived uncertainty measures,
grounded in the moment conditions, preserve the conventional GMM interpretation,
with the covariance estimated as 
\[
\hat{\boldsymbol{V}}=\frac{1}{T}\left(\hat{\boldsymbol{G}}^{\top}\boldsymbol{W}\hat{\boldsymbol{G}}\right)^{-1}\hat{\boldsymbol{G}}^{\top}\boldsymbol{W}\hat{\boldsymbol{\Sigma}}\boldsymbol{W}\hat{\boldsymbol{G}}\left(\hat{\boldsymbol{G}}^{\top}\boldsymbol{W}\hat{\boldsymbol{G}}\right)^{-1},
\]
where $\hat{\boldsymbol{\Sigma}}$ is the empirical covariance matrix
of $\boldsymbol{m}_{t}\left(\hat{\boldsymbol{\theta}}\right).$ Pointwise
confidence intervals are constructed from the corresponding diagonal
elements of $\hat{\boldsymbol{V}}$. The sample analogue $\hat{\boldsymbol{\Sigma}}$
is computed as the empirical covariance matrix of $\boldsymbol{m}_{t}\left(\hat{\boldsymbol{\theta}}\right)$,
consistent with the baseline covariance choice discussed in Section
3.2. Simultaneous confidence bands are constructed using the max-$t$
procedure of \citeauthor{MontielOlea2019} (2019, Section 2.5), applied
to the impulse-response coefficients. Since these intervals and bands
are sandwich-based, they should be understood as frequentist confidence
intervals and confidence bands centered on the regularized estimator,
rather than as posterior credible intervals. Specifically, they do
not directly account for posterior uncertainty concerning the smoothing-scale
parameters $\tau_{j}$. In the baseline analysis, the hierarchical
prior mainly functions to regularize the point estimate by determining
the degree of cross-horizon shrinkage learned from the data. 

\section{Simulation Study}

To assess the finite-sample performance of the proposed LP--IV estimator
under genuine endogeneity, we conduct a series of Monte Carlo experiments.
See Appendix A.1 for the data-generating process. We consider two
priors for the LP--IV coefficients. First, to make a direct comparison
with frequentist methods, we use a flat prior, $p\left(\boldsymbol{\theta}\right)\propto1$.
Second, to assess the practical value of regularization, we use the
roughness-penalty prior introduced in Section 3.3. We refer to the
former as QB--flat and the latter as QB--RP. We consider sample
sizes $T\in\left\{ 200,500,1000\right\} $. The hyperparameter $\rho$
ensures that the prior correlation between consecutive parameters
is approximately 0.5 and the prior correlation between the two endpoints
is approximately zero. \footnote{$\exp\left(-\sqrt{8}/4\times1\right)\approx0.493,$ $\exp\left(-\sqrt{8}/4\times3\right)\approx0.120$,
$\exp\left(-\sqrt{8}/4\times5\right)\approx0.029$, and $\exp\left(-\sqrt{8}/4\times7\right)\approx0.007$.} For each design, we generate 1,000 Monte Carlo samples. Each MCMC
run uses 25,000 iterations, with the first 5,000 discarded as burn-in.

We compare the proposed quasi-Bayesian method with a traditional benchmark:
the system-wide two-step GMM estimator. In the first stage, we use
the identity weighting matrix, which is equivalent to the two-stage
least-squares estimator, and in the second stage, we apply the weighting
matrix evaluated at the first-stage estimate. Consistent with the
covariance choice discussed in Section 3.2, the main simulation results
use the empirical covariance matrix of the stacked moment scores rather
than a HAR long-run covariance estimator. In supplementary simulations
reported in Appendix A, we also examine alternative weighting and
covariance estimators, including a block-diagonal covariance estimator
and a HAR estimator with a Bartlett kernel \citep{Newey1987}. For
the HAR estimator, the bandwidth is $B=\left\lceil 1.3T^{1/2}\right\rceil $,
following \citet{Lazarus2018}. In the exactly identified case, the
unregularized IV/GMM point estimate is invariant to the weighting
matrix. By contrast, the roughness-penalty quasi-posterior mean can
depend on the weighting matrix in finite samples because $\boldsymbol{W}$
affects the curvature of the GMM quasi-likelihood, $T\hat{\boldsymbol{G}}^{\top}\boldsymbol{W}\hat{\boldsymbol{G}}$,
and therefore the finite-sample trade-off between fitting the IV moments
and enforcing cross-horizon smoothness. Performance is evaluated by
mean bias, RMSE, mean pointwise 90\% interval length, pointwise 90\%
coverage, and simultaneous 90\% coverage across horizons.

Tables 1 and 2 present Monte Carlo results for the LD specification,
using the plain covariance estimator as the benchmark. Table 1 reports
pointwise performance measures, while Table 2 reports simultaneous
90\% coverage. The quasi-Bayesian estimator with a flat prior (QB--flat)
closely matches the classical GMM estimator across sample sizes and
horizons, due to the equivalence between the flat-prior quasi-posterior
mean and the GMM estimate. Bias remains small and diminishes with
larger samples (Panel (a) of Table 1), while RMSE decreases as T increases
and tends to rise with the horizon because sampling uncertainty builds
up (Panel (b) of Table 1). Pointwise interval lengths for GMM and
QB--flat are also almost the same (Panel (c) of Table 1), and pointwise
coverage is generally close to the nominal 90\% level (Panel (d) of
Table 1). Table 2 shows that simultaneous coverage is close to 90\%
for GMM and QB--flat. For QB--RP, coverage is conservative in smaller
samples and approaches the nominal level as the sample size increases.

The quasi-Bayesian estimator with the roughness-penalty prior (QB--RP)
smooths coefficient paths across different horizons. This regularization
introduces some bias compared to GMM and QB--flat, especially in
small samples, but it often decreases RMSE at longer horizons. The
most notable improvement occurs at $T=200$, while in larger samples,
the benefits are mainly seen at later horizons. This improvement comes
from the prior's ability to borrow information from nearby horizons
and minimize spurious high-frequency variations in the impulse response
estimate. Therefore, the quasi-Bayesian approach incorporates smoothness
directly into the estimation process rather than applying it after
estimation.

Tables A.1--A.5 in Appendix A report supplementary simulations comparing
the level and LD specifications alongside alternative weighting and
covariance estimators. The key patterns stay consistent across different
scenarios. The LD specification generally shows slightly less bias
in several cases, aligning with evidence that long-difference LP estimators
can mitigate finite-sample bias when shocks contain noise \citep{Piger2025}.
The results from the baseline and block-diagonal covariance estimators
are similar, with the block-diagonal option offering a more stable
numerical approximation when estimating the full covariance matrix
proves challenging. The HAR estimator tends to produce shorter confidence
intervals in some cases and often results in lower coverage. Overall,
QB--flat remains nearly equivalent to GMM across all specifications.
The RMSE improvements with QB--RP are most notable with the baseline
and block-diagonal estimators, especially at medium and longer horizons.
However, when using the HAR estimator, QB--RP does not always improve
RMSE and may perform worse in certain designs.

\section{Application: Renewable Energy and Electricity Prices}

This section uses the framework to analyze daily renewable-generation
and wholesale-price data. The goal is to estimate how external changes
in renewable supply influence electricity prices in the following
days.

\subsection{Data and sample construction}

The empirical analysis uses daily data for Denmark's two wholesale
electricity bidding zones: Western Denmark (DK1) and Eastern Denmark
(DK2). Market data are sourced from Open Power System Data. The boundaries
of the bidding zones are defined using Eurostat GISCO country geometries
and 2021 NUTS-2 regions: DK1 includes Syddanmark, Midtjylland, and
Nordjylland, while DK2 covers Hovedstaden and Sjælland. The dataset
spans from January 1, 2015, to September 30, 2020.

Daily electricity prices are calculated as averages of hourly prices
determined one day in advance. Wind and solar generation are measured
as daily totals, aggregated from hourly data, while electricity demand
is represented by the total daily load. Meteorological variables are
derived from ERA5 hourly single-level reanalysis data available through
the Copernicus Climate Data Store. Wind power potential is obtained
through a nonlinear transformation of hourly wind speeds at 100 meters,
calculated from zonal and meridional wind components. Solar potential
is assessed by the daily accumulated surface downward solar radiation.
Renewable potentials at the grid-cell level are aggregated to the
bidding-zone level using technology-specific installed capacity weights.
The daily wind and solar potential series are employed as instruments
for observed wind and solar generation. Weather influences, including
daily mean 2m temperature, total precipitation, and total cloud cover,
are derived from the same ERA5 data to account for weather-driven
variations in electricity demand.

\subsection{Model specification}

Let $y_{t}$ denote the daily average wholesale electricity price,
and $r_{t}^{wind}$ and $r_{t}^{solar}$ denote daily wind and solar
generation, respectively. We estimate the LD specification 
\begin{eqnarray*}
y_{t+h}-y_{t-1} & = & \gamma_{\left(h\right)}^{wind}r_{t}^{wind}+\gamma_{\left(h\right)}^{solar}r_{t}^{solar}+\eta_{\left(h\right)}+\sum_{l=1}^{L}\mu_{\left(h\right),l}\Delta y_{t-l}\\
 &  & +\varsigma_{t}\left(\boldsymbol{\alpha}_{\left(h\right)},\boldsymbol{\beta}_{\left(h\right)}\right)+\breve{\boldsymbol{\delta}}_{\left(h\right)}{}^{\top}\breve{\boldsymbol{x}}_{t}+e_{\left(h\right),t}.
\end{eqnarray*}
The coefficients $\gamma_{\left(h\right)}^{wind}$ and $\gamma_{\left(h\right)}^{solar}$
trace the dynamic price responses to wind and solar generation. Wind
and solar generation are normalized to have a mean of zero and a variance
of one, so these coefficients represent the average change in price
in response to a one-standard-deviation increase in generation. To
account for strong seasonal variation, we include a Fourier series
in calendar time, $\varsigma_{t}\left(\boldsymbol{\alpha}_{\left(h\right)},\boldsymbol{\beta}_{\left(h\right)}\right)$.
Let $d_{t}$ denote the day-of-year and $D\left(t\right)\in\left\{ 365,366\right\} $
be the number of days in the corresponding year. To handle leap years,
define the normalized within-year position $s_{t}=\left(d_{t}-1\right)/D\left(t\right)$.
The seasonal component is 
\[
\varsigma_{t}\left(\boldsymbol{\alpha}_{\left(h\right)},\boldsymbol{\beta}_{\left(h\right)}\right)=\sum_{n=1}^{N}\left[\alpha_{\left(h\right),n}\sin\left(2\pi ns_{t}\right)+\beta_{\left(h\right),n}\cos\left(2\pi ns_{t}\right)\right],
\]
where 
\[
\boldsymbol{\alpha}_{\left(h\right)}=\left(\alpha_{\left(h\right),1},...,\alpha_{\left(h\right),N}\right)^{\top},\quad\boldsymbol{\beta}_{\left(h\right)}=\left(\beta_{\left(h\right),1},...,\beta_{\left(h\right),N}\right)^{\top}.
\]
The baseline specification uses $N=4$.

The covariate vector $\breve{\boldsymbol{x}}_{t}$ includes variables
capturing demand conditions and systematic market patterns. The continuous
controls include total electricity load, daily mean 100m wind speed,
daily mean 2m temperature, total precipitation, and total cloud cover;
each is standardized to have a mean of zero and a variance of one.
Additionally, we incorporate day-of-week indicators, excluding Monday,
along with four calendar indicators: public holidays, Constitution
Day, Christmas Eve, and New Year's Eve.\footnote{Public holidays include New Year's Day, Maundy Thursday, Good Friday,
Easter Sunday, Easter Monday, Great Prayer Day, Ascension Day, Whit
Sunday / Pentecost, Whit Monday, Christmas Day, and Second Christmas
Day. }

The resulting regressor vector is 
\begin{eqnarray*}
\boldsymbol{x}_{t} & = & \left(r_{t}^{wind},r_{t}^{solar},1,\Delta y_{t-1},...,\Delta y_{t-L},\sin\left(2\pi1s_{t}\right),...,\sin\left(2\pi Ns_{t}\right),\right.\\
 &  & \quad\left.\cos\left(2\pi1s_{t}\right),...,\cos\left(2\pi Ns_{t}\right),\breve{\boldsymbol{x}}_{t}^{\top}\right)^{\top}.
\end{eqnarray*}
The corresponding coefficient vector $\boldsymbol{\theta}_{\left(h\right)}$
is defined as in Section 3.1. We set $H=L=7$ and use the roughness-penalty
prior with $\rho=4$ and $\kappa=1$. The MCMC simulation uses 50,000
draws after discarding the initial 5,000 draws as burn-in.

\subsection{Identification strategy and instrument construction}

Renewable generation might be endogenous due to potential correlation
with unobserved factors affecting electricity prices, such as demand
changes and weather patterns. To mitigate this issue, meteorological
variables are employed as instruments for renewable generation. These
instruments are derived from daily wind speed, adjusted to account
for the nonlinear link between wind speed and turbine output, along
with daily solar irradiance. Both variables serve as robust predictors
of renewable generation because of the direct physical connection
between weather conditions and energy yield \citep{Staffell2016,Pfenninger2016}.

The wind instrument is constructed from 100m wind speed. For each
grid cell $i$ and hour $s$, we define 
\[
WS_{i,s}=\sqrt{u_{i,s}^{2}+v_{i,s}^{2}},
\]
where $u_{i,s}$ and $v_{i,s}$ are the 100m zonal and meridional
wind components, respectively. Wind speed is transformed into wind-power
potential using the turbine power curve 
\[
q_{i,s}^{wind}=\begin{cases}
0, & WS_{i,s}<\varrho^{in},\\
\frac{WS_{i,s}^{3}-\left(\varrho^{in}\right)^{3}}{\left(\varrho^{rated}\right)^{3}-\left(\varrho^{in}\right)^{3}}, & \varrho^{in}\leq WS_{i,s}<\varrho^{rated},\\
1, & \varrho^{rated}\leq WS_{i,s}<\varrho^{out},\\
0, & WS_{i,s}\geq\varrho^{out}.
\end{cases}
\]
We set $\varrho^{in}=3$, $\varrho^{rated}=13$, and $\varrho^{out}=25$.
The daily grid-cell wind potential is 
\[
q_{i,t}^{wind}=\frac{1}{24}\sum_{s\in t}q_{i,s}^{wind}.
\]
We then aggregate to the bidding-zone level using installed wind-capacity
weights, 
\[
q_{t}^{wind}=\sum_{i}w_{i}^{wind}q_{i,t}^{wind},
\]
\[
w_{i}^{wind}=\frac{c_{i}^{wind}}{\sum_{j}c_{j}^{wind}}.
\]
This procedure produces a wind-potential series for bidding zones
that approximates the weather-driven part of total wind generation.
Capacity weighting is intuitive, as it reflects how meteorological
conditions relate to turbine locations. We create these weights using
renewable power-plant data from Open Power System Data, which includes
information on Danish renewable facilities and their installed capacities.
The solar instrument is constructed from surface downward solar radiation.
For each grid cell $i$ and hour $s$, let $SSR_{i,s}$ denote the
hourly accumulated surface solar radiation from ERA5. The daily grid-cell
solar potential is

\[
q_{i,t}^{solar}=\sum_{s\in t}SSR_{i,s}.
\]
We aggregate to the bidding-zone level using installed solar-capacity
weights:

\[
q_{t}^{solar}=\sum_{i}w_{i}^{solar}q_{i,t}^{solar},\quad w_{i}^{solar}=\frac{c_{i}^{solar}}{\sum_{j}c_{j}^{solar}}.
\]

The instrument vector is obtained by replacing the observed wind and
solar generation in $\boldsymbol{x}_{t}$ with the corresponding weather-driven
renewable potentials: 
\begin{eqnarray*}
\boldsymbol{z}_{t} & = & \left(q_{t}^{wind},q_{t}^{solar},1,\Delta y_{t-1},...,\Delta y_{t-L},\sin\left(2\pi1s_{t}\right),...,\sin\left(2\pi Ns_{t}\right),\right.\\
 &  & \quad\left.\cos\left(2\pi1s_{t}\right),...,\cos\left(2\pi Ns_{t}\right),\breve{\boldsymbol{x}}_{t}^{\top}\right)^{\top}.
\end{eqnarray*}
The horizon-$h$ moment condition is 
\[
\mathbb{E}\left[\left(y_{t+h}-y_{t-1}-\boldsymbol{\theta}_{\left(h\right)}^{\top}\boldsymbol{x}_{t}\right)\boldsymbol{z}_{t}\right]=\boldsymbol{0}_{J},
\]
which requires that, conditional on controls, weather-driven renewable
potential affects electricity prices only through realized renewable
generation.

The core assumption is that, once demand, weather, seasonal, and calendar
factors are held constant, changes in renewable potential influence
electricity prices solely via renewable generation. While weather
can directly affect electricity demand, this is mitigated by flexibly
controlling for variables like temperature, precipitation, cloud cover,
seasonality, and calendar effects. With this exclusion restriction,
any remaining variation in renewable potential serves as an exogenous
source of renewable supply variation. This approach aligns with empirical
research that treats renewable variability as quasi-experimental variation
in electricity markets \citep{Hirth2013}.

The instruments used in this study should be viewed as weather-state
instruments rather than weather-innovation instruments. Weather-innovation
instruments involve unexpected meteorological changes and are designed
to be orthogonal to predetermined market variables by construction.
Conversely, our instruments are based on actual wind and solar potential
levels. These factors are strong physical predictors of renewable
energy output but, due to weather persistence, may also be correlated
with lagged renewable output, historical prices, or other prior market
conditions. Thus, identification depends on a conditional weather-state
exclusion restriction: once demand, weather controls, seasonality,
calendar effects, and lagged outcomes are accounted for, the residual
variation in renewable potential influences electricity prices solely
through realized renewable generation. The placebo diagnostics in
Appendix B serve as tests of this conditional restriction's plausibility,
rather than as evidence that the instruments are purely weather innovations.

\subsection{Instrument diagnostics}

We conduct diagnostics to evaluate the IV design, with details available
in Appendix B. To assess first-stage relevance, we regress renewable
generation on renewable potential and baseline controls, reporting
coefficient estimates, the partial $R^{2}$ for the excluded instruments,
Wald-type first-stage relevance statistics, and the smallest singular
value of the matrix of renewable-potential coefficients (see Table
B.1). The lowest first-stage relevance statistic is 45.6, and the
smallest singular value is 0.336, both indicating strong relevance.
In placebo regressions, most partial $R^{2}$ values are small, except
for a maximum of 0.301, showing that current renewable potential can
predict lagged renewable generation in DK2.

Second, we perform placebo diagnostics for predetermined variables.
Specifically, we regress lagged market variables---such as prices,
total load, and renewable generation---on renewable potential and
baseline controls. We then report the coefficient estimates, partial
$R^{2}$, and Wald $p$-values (see Table B.2). The placebo partial
$R^{2}$ values are typically small, but higher values and lower Wald
$p$-values for lagged renewable generation suggest that the excluded
instruments still contain predictable weather-state components. Some
$p$-values remain small, indicating the instruments are more aligned
with weather-state instruments than purely weather-innovation instruments.
This is especially evident in DK2, where current potential predicts
lagged renewable generation, likely reflecting serial correlation
in wind and solar conditions. Consequently, Section 5.6 presents a
specification including lagged renewable potential and lagged renewable
generation. The baseline model excludes these lagged renewable-state
controls because they can reduce interpretability and may absorb part
of the persistent supply-shock variation that the IRF aims to capture.

These placebo results do not definitively disprove the empirical design,
but they suggest that the instruments maintain predictable weather-state
elements. Consequently, the estimates should be understood within
the framework of the conditional weather-state exclusion restriction
outlined earlier, rather than as outcomes driven by unforeseen weather
changes. This distinction is particularly important for DK2, where
recent renewable potential forecasts likely influence lagged renewable
generation, plausibly indicating serial correlation in wind and solar
conditions.

These diagnostics are not utilized to build the quasi-posterior. Instead,
they evaluate whether the empirical design provides enough exogenous
variation to identify the dynamic effects of wind and solar generation.
They also check whether the excluded weather-potential instruments
predict predetermined market conditions after accounting for the baseline
controls. While the estimator has a quasi-Bayesian nature, its identifying
assumptions impose restrictions on the data-generating process, which
can be tested using traditional design-based diagnostics. Developing
quasi-Bayesian tests for IV moment restrictions remains a topic for
future research.

Finally, to identify pre-existing price movements, we perform a lead-placebo
analysis. We choose the level specification because the goal of this
exercise is to determine whether renewable-potential instruments forecast
pre-existing price levels, not to measure the post-shock cumulative
price response. We regress pre-treatment leads, $y_{t-8},...,y_{t-1}$,
on the same regressors as in the baseline model. Significant responses
at these leads would suggest pre-existing price movements linked to
renewable potential; however, the absence of such responses supports,
but does not prove, the exclusion restriction. As shown in Figure
B.1, lead responses are generally near zero, and the uncertainty ranges
do not show a clear pre-trend pattern.

\subsection{Results}

Figure 1 reports the estimated IRFs for DK1 and DK2. The responses
are measured in EUR/MWh and represent a one-standard-deviation increase
in wind or solar generation. The key result is that wind generation
significantly lowers wholesale electricity prices in the short run,
while the effect of solar is smaller and less accurately estimated.

Wind generation has a negative estimated effect on both bidding zones.
In DK1, a one-standard-deviation rise in wind output reduces the daily
average electricity price by around 8 EUR/MWh initially. This effect
diminishes over three to four days, approaching zero. In DK2, the
negative effect is also observed but is smaller, about 4 EUR/MWh,
and similarly decreases over time. These findings align with the merit-order
principle: low-marginal-cost wind shifts the supply curve outward,
displacing more expensive thermal generation and lowering wholesale
prices. These results corroborate previous research on renewable energy
and electricity prices \citep{Sensfuss2008,Wuerzburg2013,Clo2015,Hirth2013}
and further show the dynamic adjustment process.

The estimated response of solar generation is subdued. In DK1, the
solar response stays near zero across different time horizons. In
DK2, the point estimates are positive at shorter horizons and decrease
over time, but the wide uncertainty intervals usually include zero.
These findings should be viewed with caution. They might indicate
that solar generation played a smaller role in Denmark during the
period studied, reflect the high intraday concentration of solar output,
or be affected by using daily average prices, which can mask within-day
merit-order effects during daylight hours. Unlike wind generation,
the daily data do not strongly suggest that solar generation causes
a significant negative price response.

The comparison between DK1 and DK2 reveals spatial differences in
how renewable generation affects prices. The wind effect is more pronounced
in DK1 than in DK2, aligning with variations in wind penetration,
interconnection, market structure, and local generation mixes. This
variation is important for policy, as the price effects of expanding
renewables can vary across bidding zones based on renewable capacity,
transmission limitations, and the presence of flexible resources.

Overall, the findings indicate that integrating renewable energy mainly
lowers wholesale prices through wind power, especially over short-term
periods. Prices respond to weather-related wind supply shocks within
a few days, though not instantaneously. This has important implications
for balancing markets, storage investments, demand response, and managing
short-term risks. Additionally, the results demonstrate the value
of the proposed quasi-Bayesian LP--IV method: it generates smooth
impulse responses that are easier to interpret economically, while
also providing uncertainty estimates through both pointwise and simultaneous
bands.

\subsection{Robustness}

To evaluate robustness, we explore several alternative specifications.
Figure C.1 shows estimates based on the level specification instead
of the LD specification. Figures C.2 and C.3 modify the lag length
and seasonal flexibility by setting $L=14$ and $N=8$, respectively.
Figures C.4 and C.5 adjust the tuning parameter by setting $\rho=2$
and $\rho=8$. Figures C.6 and C.7 change the prior scale to $\kappa=0.1$
and $\kappa=10$. Guided by the weather-state nature of the instruments
and the diagnostics for predetermined variables in Section 5.4, Figure
C.8 extends the baseline specification by including lagged wind and
solar generation, as well as lagged wind and solar potential from
$t-1$ to $t-7$. Finally, Figure C.9 presents estimates under the
flat prior.

The key findings remain consistent across different exercises. Wind
responses consistently show a negative effect in both DK1 and DK2,
gradually approaching zero. While the magnitude of responses varies
depending on the specifications, the overall pattern remains the same:
wind generation leads to a short-term decrease in wholesale electricity
prices, with the most significant effect occurring at or shortly after
impact. Solar responses follow the same overall trend as in the baseline
but are smaller and less precisely measured than wind responses. Robustness
checks do not indicate a consistent negative daily price response
to solar generation.

The lag-augmented specification tackles the concern that baseline
estimates might reflect serial correlation in weather or renewable
generation. The estimates remain largely similar to the original,
particularly for short-term wind responses. This suggests that the
baseline results are primarily influenced by current weather-driven
renewable supply fluctuations rather than ongoing renewable condition
persistence. We do not adopt the lag-augmented model as our standard
baseline because including lagged renewable state controls could absorb
part of the dynamic adjustment that the IRF aims to capture.

The comparison between the roughness-penalty prior and the flat prior
reveals that regularization primarily stabilizes the estimated response
path. As anticipated, the flat prior results in less smooth estimates,
yet the overall economic interpretation stays consistent. This aligns
with the simulation findings in Section 4, where the roughness-penalty
prior decreases sampling variability without altering the key implications
of the IV estimates.

Overall, the robustness check reinforces the main results from Figure
1: wind generation consistently lowers prices in the short run in
both Danish bidding zones, while solar effects are weaker and less
precisely measured.

\section{Conclusion}

This paper presents a quasi-Bayesian LP--IV approach for estimating
dynamic causal effects. The method addresses a common challenge in
LP--IV applications: horizon-specific estimates can be flexible but
often noisy, especially at longer horizons and with high-frequency
data. To mitigate this, the approach treats the IRF as a series of
related parameters and employs a roughness-penalty prior to smooth
out variations across neighboring horizons. Since inference relies
on moment conditions rather than a full likelihood, the framework
preserves the interpretability of IV estimation while enabling structured
regularization of the entire response path.

The method's theoretical and computational framework is straightforward.
Building on \citet{Chernozhukov2003}, the quasi-posterior is derived
from the GMM objective function. Under typical regularity conditions,
the quasi-posterior mean aligns asymptotically with the traditional
GMM estimator, ensuring that the proposed estimator maintains the
primary properties of standard LP--IV. Additionally, the roughness-penalty
prior enhances finite-sample stability by sharing information across
different horizons. The simulation of the quasi-posterior is carried
out using a Gibbs sampler, which takes advantage of the conditional
Gaussian structure of the quasi-posterior and the structured precision
matrix created by the prior.

The simulation results indicate that the proposed estimator performs
effectively in finite samples. Using the flat prior, the quasi-Bayesian
estimator closely aligns with the classical GMM, reinforcing the connection
between the quasi-posterior and traditional moment-based methods.
When employing the roughness-penalty prior, the estimator applies
moderate shrinkage, which often leads to a lower RMSE, especially
at medium and longer horizons. This is because unrestricted LP--IV
estimates tend to be more influenced by sampling noise at these horizons.
Overall, these findings support applying cross-horizon regularization
when the goal is to obtain a smooth impulse-response function.

The empirical study examines how renewable generation affects wholesale
electricity prices in Denmark over time. By using weather-driven renewable
potential as instruments for wind and solar output, the findings reveal
that wind generation significantly lowers prices in both Danish bidding
zones in the short term. The effect is more pronounced in Western
Denmark and diminishes over subsequent days, suggesting that the merit-order
effect is most relevant at short horizons. Conversely, solar effects
are smaller and less precisely estimated with daily data. These results
align with existing merit-order literature and advance it by analyzing
the complete dynamic price response to renewable supply shocks. Robustness
tests confirm that the main conclusions remain consistent across various
LP models, lag structures, seasonal controls, prior hyperparameters,
lag-augmented renewable controls, and placebo exercises.

The findings have implications for both econometric practice and energy-market
policy. Methodologically, the study demonstrates that quasi-Bayesian
regularization is effective when the target parameter is a path indexed
by a horizon, rather than a single scalar coefficient. Substantively,
the empirical results indicate that renewable energy integration influences
not only current wholesale electricity prices but also the short-term
price dynamics. This is significant for designing balancing markets,
storage incentives, demand-response programs, and risk management
strategies. As electricity systems increasingly include larger proportions
of variable renewable energy, methods capable of reliably estimating
dynamic effects amid endogeneity and high-frequency volatility will
gain greater importance.

Several avenues remain for future research. One involves extending
the framework to overidentified LP--IV systems and creating quasi-Bayesian
procedures for testing overidentifying restrictions and assessing
instrument validity within the same moment-based approach. Another
potential is to develop procedures robust to weak instruments and
methods for instrument selection in quasi-Bayesian LP--IV \citep{Goh2022}.
Additionally, enhancing scalability with respect to sample size, projection
horizon, and the number of explanatory variables is crucial. These
advancements would significantly expand the applicability of quasi-Bayesian
LP--IV methods in applied econometrics.

\paragraph{Funding Statement}

This research received no specific grant from any funding agency in
the public, commercial, or not-for-profit sectors.

\paragraph{Conflict of Interest Statement}

The author declares no conflict of interest.

\bibliographystyle{apalike2}
\bibliography{reference}

\clearpage{}

\begin{table}
\caption{Simulation result (1): Pointwise performance measures}

\medskip{}

\centering{}%
\begin{tabular}{llr@{\extracolsep{0pt}.}lr@{\extracolsep{0pt}.}lr@{\extracolsep{0pt}.}lr@{\extracolsep{0pt}.}lr@{\extracolsep{0pt}.}lr@{\extracolsep{0pt}.}lr@{\extracolsep{0pt}.}lr@{\extracolsep{0pt}.}l}
\hline 
\multirow{2}{*}{$T$} & \multirow{2}{*}{Method} & \multicolumn{16}{c}{$h$}\tabularnewline
\cline{3-18}
 &  & \multicolumn{2}{c}{0} & \multicolumn{2}{c}{1} & \multicolumn{2}{c}{2} & \multicolumn{2}{c}{3} & \multicolumn{2}{c}{4} & \multicolumn{2}{c}{5} & \multicolumn{2}{c}{6} & \multicolumn{2}{c}{7}\tabularnewline
\hline 
\multicolumn{18}{l}{(a) Mean bias}\tabularnewline
\hline 
\multirow{3}{*}{200} & GMM & 0&002 & 0&002 & 0&000 & 0&000 & 0&003 & 0&002 & 0&001 & 0&000\tabularnewline
 & QB--flat & 0&002 & 0&002 & 0&000 & 0&000 & 0&003 & 0&003 & 0&001 & 0&000\tabularnewline
 & QB--RP & -0&023 & -0&045 & -0&054 & -0&052 & -0&044 & -0&038 & -0&035 & -0&030\tabularnewline
\hline 
\multirow{3}{*}{500} & GMM & 0&001 & 0&001 & -0&001 & -0&001 & 0&000 & -0&003 & -0&003 & -0&004\tabularnewline
 & QB--flat & 0&001 & 0&001 & -0&001 & -0&001 & 0&000 & -0&003 & -0&003 & -0&004\tabularnewline
 & QB--RP & -0&021 & -0&039 & -0&045 & -0&043 & -0&038 & -0&035 & -0&031 & -0&028\tabularnewline
\hline 
\multirow{3}{*}{1000} & GMM & -0&001 & -0&003 & -0&004 & -0&004 & -0&005 & -0&004 & -0&002 & 0&001\tabularnewline
 & QB--flat & -0&001 & -0&003 & -0&004 & -0&004 & -0&005 & -0&004 & -0&002 & 0&001\tabularnewline
 & QB--RP & -0&017 & -0&031 & -0&036 & -0&035 & -0&032 & -0&028 & -0&023 & -0&017\tabularnewline
\hline 
\multicolumn{18}{l}{(b) RMSE}\tabularnewline
\hline 
\multirow{3}{*}{200} & GMM & 0&089 & 0&118 & 0&129 & 0&140 & 0&149 & 0&148 & 0&148 & 0&151\tabularnewline
 & QB--flat & 0&089 & 0&118 & 0&129 & 0&140 & 0&149 & 0&148 & 0&148 & 0&151\tabularnewline
 & QB--RP & 0&086 & 0&110 & 0&115 & 0&118 & 0&116 & 0&114 & 0&112 & 0&117\tabularnewline
\hline 
\multirow{3}{*}{500} & GMM & 0&057 & 0&072 & 0&079 & 0&087 & 0&088 & 0&090 & 0&094 & 0&094\tabularnewline
 & QB--flat & 0&057 & 0&072 & 0&079 & 0&087 & 0&088 & 0&090 & 0&094 & 0&094\tabularnewline
 & QB--RP & 0&060 & 0&078 & 0&083 & 0&083 & 0&078 & 0&076 & 0&078 & 0&078\tabularnewline
\hline 
\multirow{3}{*}{1000} & GMM & 0&040 & 0&052 & 0&056 & 0&061 & 0&062 & 0&063 & 0&063 & 0&065\tabularnewline
 & QB--flat & 0&040 & 0&052 & 0&056 & 0&061 & 0&062 & 0&063 & 0&063 & 0&065\tabularnewline
 & QB--RP & 0&044 & 0&060 & 0&064 & 0&065 & 0&063 & 0&060 & 0&057 & 0&058\tabularnewline
\hline 
\multicolumn{18}{l}{(c) Mean pointwise 90\% interval length}\tabularnewline
\hline 
\multirow{3}{*}{200} & GMM & 0&290 & 0&374 & 0&420 & 0&449 & 0&464 & 0&473 & 0&480 & 0&485\tabularnewline
 & QB--flat & 0&290 & 0&374 & 0&420 & 0&449 & 0&464 & 0&473 & 0&480 & 0&485\tabularnewline
 & QB--RP & 0&293 & 0&379 & 0&426 & 0&455 & 0&471 & 0&480 & 0&487 & 0&491\tabularnewline
\hline 
\multirow{3}{*}{500} & GMM & 0&184 & 0&237 & 0&265 & 0&283 & 0&294 & 0&300 & 0&305 & 0&308\tabularnewline
 & QB--flat & 0&184 & 0&237 & 0&265 & 0&283 & 0&294 & 0&300 & 0&305 & 0&308\tabularnewline
 & QB--RP & 0&185 & 0&239 & 0&267 & 0&284 & 0&295 & 0&302 & 0&306 & 0&309\tabularnewline
\hline 
\multirow{3}{*}{1000} & GMM & 0&130 & 0&167 & 0&188 & 0&199 & 0&207 & 0&211 & 0&214 & 0&216\tabularnewline
 & QB--flat & 0&130 & 0&167 & 0&188 & 0&199 & 0&207 & 0&211 & 0&214 & 0&216\tabularnewline
 & QB--RP & 0&131 & 0&168 & 0&188 & 0&200 & 0&207 & 0&211 & 0&214 & 0&216\tabularnewline
\hline 
\multicolumn{18}{l}{(d) Pointwise 90\% coverage}\tabularnewline
\hline 
\multirow{3}{*}{200} & GMM & 0&897 & 0&884 & 0&899 & 0&883 & 0&870 & 0&887 & 0&899 & 0&900\tabularnewline
 & QB--flat & 0&898 & 0&882 & 0&896 & 0&883 & 0&868 & 0&887 & 0&900 & 0&901\tabularnewline
 & QB--RP & 0&918 & 0&920 & 0&939 & 0&950 & 0&957 & 0&965 & 0&962 & 0&956\tabularnewline
\hline 
\multirow{3}{*}{500} & GMM & 0&894 & 0&897 & 0&902 & 0&886 & 0&901 & 0&919 & 0&894 & 0&905\tabularnewline
 & QB--flat & 0&894 & 0&897 & 0&902 & 0&886 & 0&900 & 0&918 & 0&894 & 0&905\tabularnewline
 & QB--RP & 0&865 & 0&875 & 0&912 & 0&919 & 0&949 & 0&963 & 0&946 & 0&950\tabularnewline
\hline 
\multirow{3}{*}{1000} & GMM & 0&898 & 0&901 & 0&908 & 0&899 & 0&900 & 0&896 & 0&909 & 0&894\tabularnewline
 & QB--flat & 0&900 & 0&902 & 0&908 & 0&899 & 0&900 & 0&897 & 0&908 & 0&894\tabularnewline
 & QB--RP & 0&866 & 0&833 & 0&853 & 0&873 & 0&899 & 0&926 & 0&942 & 0&938\tabularnewline
\hline 
\end{tabular}
\end{table}

\clearpage{}

\begin{table}
\caption{Simulation result (2): Simultaneous 90\% coverage}

\medskip{}

\centering{}%
\begin{tabular}{lr@{\extracolsep{0pt}.}lr@{\extracolsep{0pt}.}lr@{\extracolsep{0pt}.}l}
\hline 
\multirow{1}{*}{Method} & \multicolumn{6}{c}{$T$}\tabularnewline
 & \multicolumn{2}{c}{200} & \multicolumn{2}{c}{500} & \multicolumn{2}{c}{1000}\tabularnewline
\hline 
GMM & 0&887 & 0&902 & 0&897\tabularnewline
QB--flat & 0&888 & 0&900 & 0&899\tabularnewline
QB--RP & 0&961 & 0&943 & 0&900\tabularnewline
\hline 
\end{tabular}
\end{table}

\clearpage{}

\begin{figure}
\caption{Estimated IRF}
\medskip{}

\begin{centering}
\includegraphics[scale=0.5]{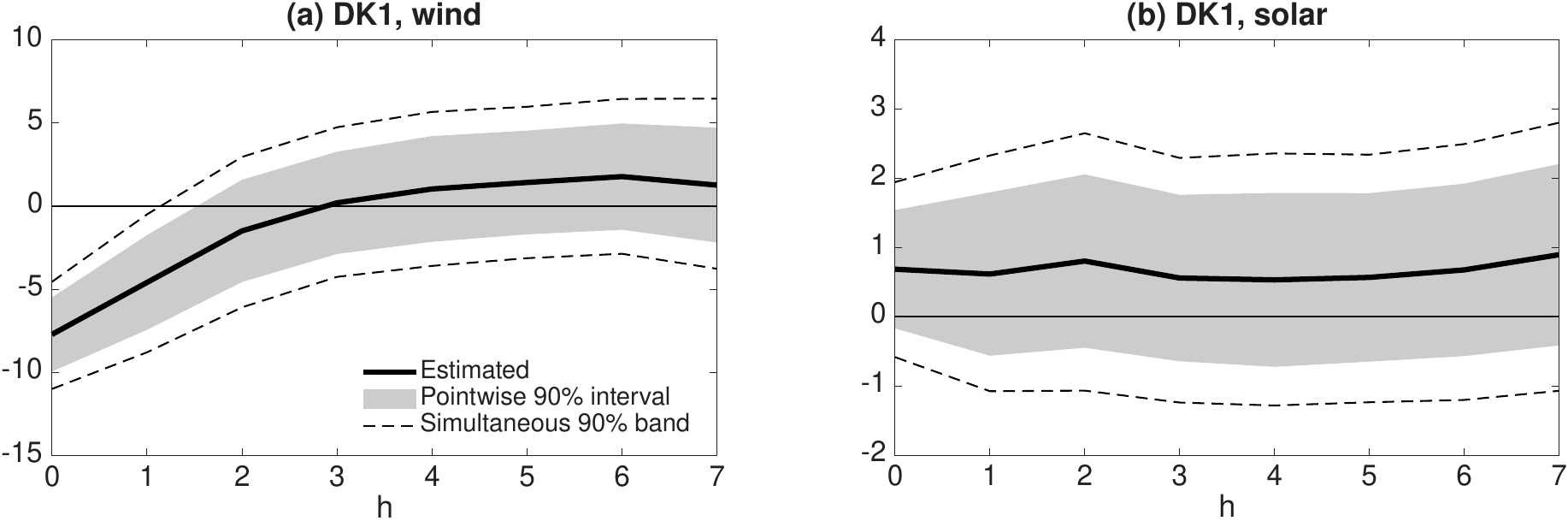}
\par\end{centering}
\medskip{}

\centering{}\includegraphics[scale=0.5]{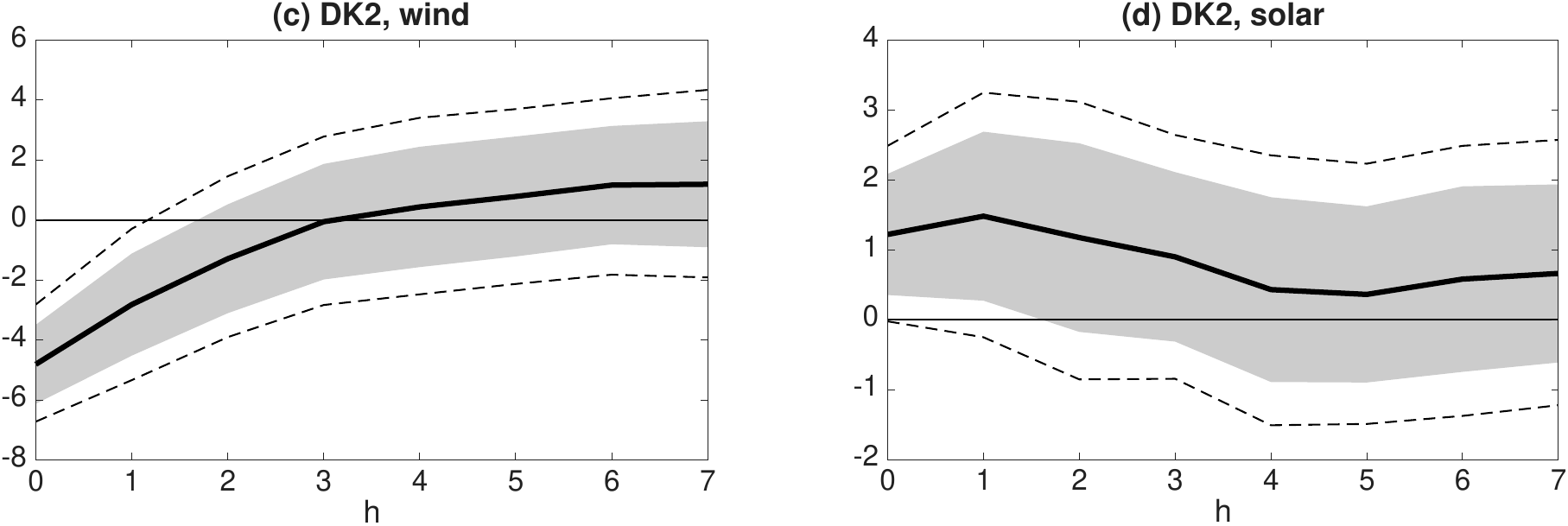}
\end{figure}

\clearpage{}



\section*{Supplementary material}

Figure A.1: IRF to be inferred

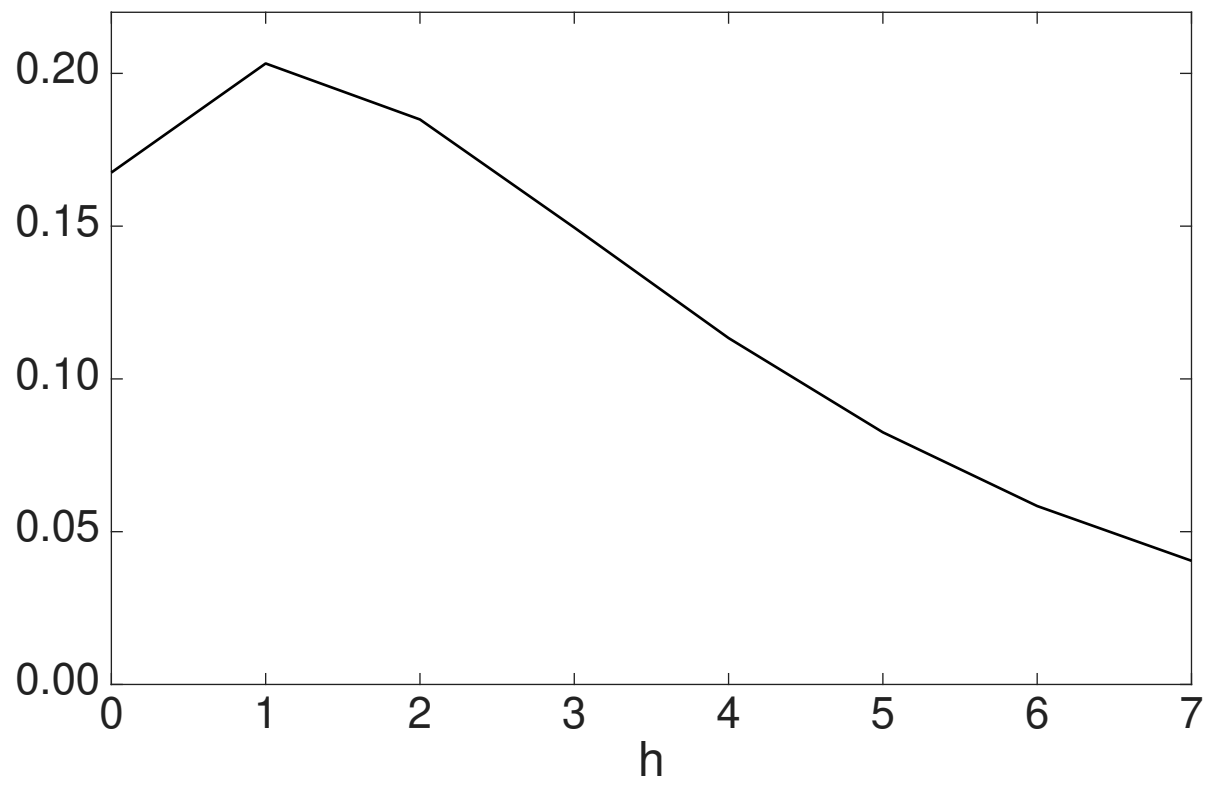


# Appendix A: Simulation Study

## A.1 Data-generating process

We generate an observed treatment that correlates with the structural outcome disturbance. Let $\varepsilon_t^{treat}$ denote the structural treatment shock, $\varepsilon_t^{\#}$ an outcome disturbance, and $\varepsilon_t^{\flat}$ an additional market-state shock. These variables are independently standard normally distributed. The observed treatment is

$$r_t = \varepsilon_t^{treat} + \lambda\varepsilon_t^{\#} + \sigma_r \epsilon_t^r, \quad \epsilon_t^r \sim \mathcal{N}(0,1).$$

When $\lambda \neq 0$, the treatment is endogenous because it is correlated with the disturbance $\varepsilon_t^{\#}$, which also affects the outcome. The excluded instrument is

$$q_t = \varepsilon_t^{treat} + \sigma_z \epsilon_t^z, \quad \epsilon_t^z \sim \mathcal{N}(0,1),$$

which is relevant for $r_t$ but independent of the outcome disturbance. The outcome is generated as

$$y_t = \sum_{l=0}^{L} \gamma_l^{IRF} \varepsilon_{t-l}^{treat} + \sum_{l=0}^{L} \gamma_l^{\sharp} \varepsilon_{t-l}^{\#} + \sum_{l=0}^{L} \gamma_l^{\flat} \varepsilon_{t-l}^{\flat}.$$

The true impulse response is $\gamma_h^{IRF}$, and the LP–IV estimator uses $q_t$ as an instrument for $r_t$. We set $H = L = 7$. The target impulse response is hump-shaped and normalized to sum to one (Figure A.1):

$$\gamma_h^{IRF} = \frac{(h+1)\exp(0.5(1-h))}{\sum_{j=0}^{H}(j+1)\exp(0.5(1-j))}.$$

The regressor vector is specified for the level and LD specifications, respectively, as

$$\boldsymbol{x}_t = (r_t, 1, y_{t-1}, ..., y_{t-L})^\top, \quad \boldsymbol{x}_t = (r_t, 1, \Delta y_{t-1}, ..., \Delta y_{t-L})^\top.$$

The nuisance coefficients are $\gamma_l^{\sharp} = 0.7^l$ and $\gamma_l^{\flat} = 0.5^{l+1}$. We fix $\lambda = 0.5$, $\sigma_r = 0.5$, and $\sigma_z = 0.5$.

## A.2 Supplementary simulation results

Table A.1: Simulation result (1): Mean bias

| $T$ | Spec. | Cov. | Method | Mean bias | | | | | | | |
|---|---|---|---|---|---|---|---|---|---|---|---|
| | | | | $h$ | | | | | | | |
| | | | | 0 | 1 | 2 | 3 | 4 | 5 | 6 | 7 |
| 200 | Level | Plain | GMM | -0.001 | -0.001 | -0.002 | -0.007 | -0.009 | -0.007 | -0.002 | -0.008 |
| | | | QB–flat | -0.001 | -0.001 | -0.002 | -0.007 | -0.009 | -0.007 | -0.003 | -0.008 |
| | | | QB–RP | -0.016 | -0.037 | -0.043 | -0.042 | -0.037 | -0.028 | -0.019 | -0.017 |
| | | Blkdiag | GMM | -0.001 | -0.001 | -0.002 | -0.007 | -0.009 | -0.007 | -0.002 | -0.008 |
| | | | QB–flat | -0.001 | -0.001 | -0.002 | -0.007 | -0.009 | -0.007 | -0.003 | -0.008 |
| | | | QB–RP | -0.005 | -0.019 | -0.019 | -0.018 | -0.016 | -0.012 | -0.007 | -0.007 |
| | | HAR | GMM | -0.001 | -0.001 | -0.002 | -0.007 | -0.009 | -0.007 | -0.002 | -0.008 |
| | | | QB–flat | -0.001 | -0.001 | -0.002 | -0.006 | -0.009 | -0.007 | -0.002 | -0.008 |
| | | | QB–RP | -0.013 | -0.030 | -0.033 | -0.034 | -0.028 | -0.022 | -0.013 | -0.014 |
| | LD | Plain | GMM | -0.004 | -0.008 | -0.008 | -0.008 | -0.004 | -0.004 | -0.004 | -0.006 |
| | | | QB–flat | -0.004 | -0.008 | -0.008 | -0.008 | -0.004 | -0.004 | -0.004 | -0.006 |
| | | | QB–RP | -0.027 | -0.053 | -0.059 | -0.057 | -0.048 | -0.041 | -0.038 | -0.035 |
| | | Blkdiag | GMM | -0.004 | -0.008 | -0.008 | -0.008 | -0.004 | -0.004 | -0.004 | -0.006 |
| | | | QB–flat | -0.004 | -0.008 | -0.008 | -0.008 | -0.004 | -0.004 | -0.004 | -0.006 |
| | | | QB–RP | -0.011 | -0.026 | -0.025 | -0.020 | -0.013 | -0.010 | -0.009 | -0.006 |
| | | HAR | GMM | -0.004 | -0.008 | -0.008 | -0.008 | -0.004 | -0.004 | -0.004 | -0.006 |
| | | | QB–flat | -0.004 | -0.008 | -0.008 | -0.008 | -0.004 | -0.004 | -0.004 | -0.006 |
| | | | QB–RP | -0.028 | -0.055 | -0.060 | -0.055 | -0.043 | -0.033 | -0.024 | -0.018 |
| 500 | Level | Plain | GMM | -0.004 | -0.005 | -0.006 | -0.005 | -0.005 | -0.006 | -0.003 | -0.003 |
| | | | QB–flat | -0.004 | -0.005 | -0.006 | -0.005 | -0.005 | -0.006 | -0.003 | -0.003 |
| | | | QB–RP | -0.018 | -0.034 | -0.039 | -0.035 | -0.030 | -0.024 | -0.016 | -0.012 |
| | | Blkdiag | GMM | -0.004 | -0.005 | -0.006 | -0.005 | -0.005 | -0.006 | -0.003 | -0.003 |
| | | | QB–flat | -0.004 | -0.005 | -0.006 | -0.005 | -0.005 | -0.006 | -0.002 | -0.003 |
| | | | QB–RP | -0.007 | -0.018 | -0.018 | -0.014 | -0.011 | -0.009 | -0.005 | -0.003 |
| | | HAR | GMM | -0.004 | -0.005 | -0.006 | -0.005 | -0.005 | -0.006 | -0.003 | -0.003 |
| | | | QB–flat | -0.004 | -0.005 | -0.006 | -0.005 | -0.005 | -0.006 | -0.002 | -0.003 |
| | | | QB–RP | -0.014 | -0.026 | -0.031 | -0.027 | -0.024 | -0.021 | -0.013 | -0.010 |
| | LD | Plain | GMM | 0.001 | -0.001 | -0.002 | -0.001 | -0.002 | 0.000 | 0.000 | -0.001 |
| | | | QB–flat | 0.001 | -0.001 | -0.002 | -0.001 | -0.002 | 0.000 | 0.000 | -0.001 |
| | | | QB–RP | -0.020 | -0.039 | -0.045 | -0.043 | -0.039 | -0.033 | -0.028 | -0.024 |
| | | Blkdiag | GMM | 0.001 | -0.001 | -0.002 | -0.001 | -0.002 | 0.000 | 0.000 | -0.001 |
| | | | QB–flat | 0.001 | -0.001 | -0.002 | -0.001 | -0.002 | 0.000 | 0.000 | -0.001 |
| | | | QB–RP | -0.004 | -0.015 | -0.014 | -0.011 | -0.008 | -0.005 | -0.003 | -0.001 |
| | | HAR | GMM | 0.001 | -0.001 | -0.002 | -0.001 | -0.002 | 0.000 | 0.000 | -0.001 |
| | | | QB–flat | 0.001 | -0.001 | -0.002 | -0.001 | -0.002 | 0.000 | 0.000 | -0.001 |
| | | | QB–RP | -0.022 | -0.042 | -0.046 | -0.043 | -0.037 | -0.028 | -0.019 | -0.011 |
| 1000 | Level | Plain | GMM | 0.000 | -0.001 | -0.005 | 0.000 | -0.001 | -0.001 | 0.000 | 0.003 |
| | | | QB–flat | 0.000 | -0.001 | -0.005 | 0.000 | -0.001 | -0.001 | 0.000 | 0.003 |
| | | | QB–RP | -0.010 | -0.023 | -0.028 | -0.022 | -0.019 | -0.015 | -0.010 | -0.005 |
| | | Blkdiag | GMM | 0.000 | -0.001 | -0.005 | 0.000 | -0.001 | -0.001 | 0.000 | 0.003 |
| | | | QB–flat | 0.000 | -0.001 | -0.005 | 0.000 | -0.001 | -0.001 | 0.000 | 0.003 |
| | | | QB–RP | -0.002 | -0.010 | -0.012 | -0.006 | -0.005 | -0.004 | -0.001 | 0.002 |
| | | HAR | GMM | 0.000 | -0.001 | -0.005 | 0.000 | -0.001 | -0.001 | 0.000 | 0.003 |
| | | | QB–flat | 0.000 | -0.001 | -0.005 | 0.000 | -0.001 | -0.001 | 0.000 | 0.003 |
| | | | QB–RP | -0.010 | -0.020 | -0.026 | -0.019 | -0.017 | -0.014 | -0.009 | -0.005 |
| | LD | Plain | GMM | -0.002 | 0.000 | 0.000 | -0.001 | -0.002 | -0.002 | 0.000 | -0.001 |
| | | | QB–flat | -0.002 | 0.000 | 0.000 | -0.001 | -0.002 | -0.002 | 0.000 | -0.001 |
| | | | QB–RP | -0.017 | -0.028 | -0.032 | -0.032 | -0.029 | -0.026 | -0.020 | -0.018 |
| | | Blkdiag | GMM | -0.002 | 0.000 | 0.000 | -0.001 | -0.002 | -0.002 | 0.000 | -0.001 |
| | | | QB–flat | -0.002 | 0.000 | 0.000 | -0.001 | -0.002 | -0.002 | 0.000 | -0.001 |
| | | | QB–RP | -0.005 | -0.010 | -0.009 | -0.008 | -0.006 | -0.005 | -0.002 | -0.001 |
| | | HAR | GMM | -0.002 | 0.000 | 0.000 | -0.001 | -0.002 | -0.002 | 0.000 | -0.001 |
| | | | QB–flat | -0.002 | 0.000 | 0.000 | -0.001 | -0.002 | -0.002 | 0.000 | -0.001 |
| | | | QB–RP | -0.019 | -0.031 | -0.035 | -0.033 | -0.029 | -0.023 | -0.015 | -0.011 |

Table A.2: Simulation result (2): RMSE

| $T$ | Spec. | Cov. | Method | RMSE | | | | | | | |
|---|---|---|---|---|---|---|---|---|---|---|---|
| | | | | $h$ | | | | | | | |
| | | | | 0 | 1 | 2 | 3 | 4 | 5 | 6 | 7 |
| 200 | Level | Plain | GMM | 0.088 | 0.109 | 0.116 | 0.118 | 0.123 | 0.125 | 0.125 | 0.125 |
| | | | QB–flat | 0.088 | 0.109 | 0.116 | 0.118 | 0.123 | 0.125 | 0.125 | 0.125 |
| | | | QB–RP | 0.083 | 0.101 | 0.105 | 0.102 | 0.101 | 0.098 | 0.098 | 0.102 |
| | | Blkdiag | GMM | 0.088 | 0.109 | 0.116 | 0.118 | 0.123 | 0.125 | 0.125 | 0.125 |
| | | | QB–flat | 0.088 | 0.109 | 0.116 | 0.118 | 0.123 | 0.125 | 0.125 | 0.125 |
| | | | QB–RP | 0.082 | 0.098 | 0.103 | 0.103 | 0.104 | 0.105 | 0.106 | 0.108 |
| | | HAR | GMM | 0.088 | 0.109 | 0.116 | 0.118 | 0.123 | 0.125 | 0.125 | 0.125 |
| | | | QB–flat | 0.088 | 0.109 | 0.116 | 0.118 | 0.123 | 0.125 | 0.125 | 0.125 |
| | | | QB–RP | 0.099 | 0.118 | 0.123 | 0.124 | 0.121 | 0.122 | 0.125 | 0.125 |
| | LD | Plain | GMM | 0.090 | 0.123 | 0.133 | 0.143 | 0.150 | 0.146 | 0.148 | 0.148 |
| | | | QB–flat | 0.090 | 0.123 | 0.133 | 0.143 | 0.150 | 0.146 | 0.148 | 0.148 |
| | | | QB–RP | 0.088 | 0.116 | 0.119 | 0.121 | 0.118 | 0.110 | 0.111 | 0.116 |
| | | Blkdiag | GMM | 0.090 | 0.123 | 0.133 | 0.143 | 0.150 | 0.146 | 0.148 | 0.148 |
| | | | QB–flat | 0.090 | 0.123 | 0.133 | 0.143 | 0.150 | 0.146 | 0.148 | 0.148 |
| | | | QB–RP | 0.086 | 0.113 | 0.120 | 0.125 | 0.129 | 0.125 | 0.125 | 0.127 |
| | | HAR | GMM | 0.090 | 0.123 | 0.133 | 0.143 | 0.150 | 0.146 | 0.148 | 0.148 |
| | | | QB–flat | 0.090 | 0.123 | 0.133 | 0.143 | 0.150 | 0.146 | 0.148 | 0.148 |
| | | | QB–RP | 0.102 | 0.136 | 0.146 | 0.150 | 0.151 | 0.147 | 0.149 | 0.158 |
| 500 | Level | Plain | GMM | 0.055 | 0.067 | 0.073 | 0.076 | 0.078 | 0.077 | 0.078 | 0.076 |
| | | | QB–flat | 0.055 | 0.067 | 0.073 | 0.076 | 0.078 | 0.077 | 0.078 | 0.076 |
| | | | QB–RP | 0.057 | 0.073 | 0.077 | 0.075 | 0.072 | 0.068 | 0.066 | 0.066 |
| | | Blkdiag | GMM | 0.055 | 0.067 | 0.073 | 0.076 | 0.078 | 0.077 | 0.078 | 0.076 |
| | | | QB–flat | 0.055 | 0.067 | 0.073 | 0.076 | 0.078 | 0.077 | 0.078 | 0.076 |
| | | | QB–RP | 0.053 | 0.066 | 0.070 | 0.070 | 0.070 | 0.069 | 0.069 | 0.069 |
| | | HAR | GMM | 0.055 | 0.067 | 0.073 | 0.076 | 0.078 | 0.077 | 0.078 | 0.076 |
| | | | QB–flat | 0.055 | 0.067 | 0.073 | 0.076 | 0.078 | 0.077 | 0.078 | 0.076 |
| | | | QB–RP | 0.062 | 0.080 | 0.083 | 0.084 | 0.083 | 0.082 | 0.079 | 0.080 |
| | LD | Plain | GMM | 0.057 | 0.074 | 0.083 | 0.087 | 0.089 | 0.090 | 0.093 | 0.090 |
| | | | QB–flat | 0.057 | 0.074 | 0.083 | 0.086 | 0.089 | 0.090 | 0.093 | 0.090 |
| | | | QB–RP | 0.060 | 0.080 | 0.085 | 0.083 | 0.080 | 0.077 | 0.077 | 0.075 |
| | | Blkdiag | GMM | 0.057 | 0.074 | 0.083 | 0.087 | 0.089 | 0.090 | 0.093 | 0.090 |
| | | | QB–flat | 0.057 | 0.074 | 0.083 | 0.086 | 0.089 | 0.090 | 0.093 | 0.090 |
| | | | QB–RP | 0.056 | 0.072 | 0.078 | 0.080 | 0.080 | 0.080 | 0.082 | 0.080 |
| | | HAR | GMM | 0.057 | 0.074 | 0.083 | 0.087 | 0.089 | 0.090 | 0.093 | 0.090 |
| | | | QB–flat | 0.057 | 0.074 | 0.083 | 0.087 | 0.089 | 0.090 | 0.093 | 0.090 |
| | | | QB–RP | 0.066 | 0.088 | 0.094 | 0.094 | 0.094 | 0.092 | 0.095 | 0.098 |
| 1000 | Level | Plain | GMM | 0.038 | 0.049 | 0.053 | 0.052 | 0.053 | 0.054 | 0.054 | 0.055 |
| | | | QB–flat | 0.038 | 0.049 | 0.053 | 0.053 | 0.053 | 0.054 | 0.054 | 0.055 |
| | | | QB–RP | 0.040 | 0.055 | 0.059 | 0.054 | 0.052 | 0.051 | 0.050 | 0.051 |
| | | Blkdiag | GMM | 0.038 | 0.049 | 0.053 | 0.052 | 0.053 | 0.054 | 0.054 | 0.055 |
| | | | QB–flat | 0.038 | 0.049 | 0.053 | 0.052 | 0.053 | 0.054 | 0.054 | 0.055 |
| | | | QB–RP | 0.037 | 0.049 | 0.052 | 0.050 | 0.049 | 0.050 | 0.050 | 0.052 |
| | | HAR | GMM | 0.038 | 0.049 | 0.053 | 0.052 | 0.053 | 0.054 | 0.054 | 0.055 |
| | | | QB–flat | 0.038 | 0.049 | 0.053 | 0.052 | 0.053 | 0.054 | 0.054 | 0.055 |
| | | | QB–RP | 0.046 | 0.060 | 0.064 | 0.060 | 0.059 | 0.059 | 0.058 | 0.059 |
| | LD | Plain | GMM | 0.040 | 0.052 | 0.056 | 0.061 | 0.062 | 0.064 | 0.065 | 0.064 |
| | | | QB–flat | 0.040 | 0.052 | 0.056 | 0.061 | 0.062 | 0.064 | 0.065 | 0.064 |
| | | | QB–RP | 0.044 | 0.059 | 0.063 | 0.064 | 0.062 | 0.060 | 0.058 | 0.058 |
| | | Blkdiag | GMM | 0.040 | 0.052 | 0.056 | 0.061 | 0.062 | 0.064 | 0.065 | 0.064 |
| | | | QB–flat | 0.040 | 0.052 | 0.056 | 0.061 | 0.062 | 0.064 | 0.065 | 0.064 |
| | | | QB–RP | 0.040 | 0.052 | 0.055 | 0.058 | 0.058 | 0.059 | 0.059 | 0.059 |
| | | HAR | GMM | 0.040 | 0.052 | 0.056 | 0.061 | 0.062 | 0.064 | 0.065 | 0.064 |
| | | | QB–flat | 0.040 | 0.052 | 0.056 | 0.061 | 0.062 | 0.064 | 0.065 | 0.064 |
| | | | QB–RP | 0.047 | 0.062 | 0.067 | 0.068 | 0.065 | 0.063 | 0.061 | 0.062 |

Table A.3: Simulation result (3): Mean pointwise 90% interval length

| $T$ | Spec. | Cov. | Method | 0 | 1 | 2 | 3 | 4 | 5 | 6 | 7 |
|---|---|---|---|---|---|---|---|---|---|---|---|
| | | | | Mean pointwise 90% interval length, $h$ | | | | | | | |
| 200 | Level | Plain | GMM | 0.276 | 0.342 | 0.373 | 0.389 | 0.395 | 0.399 | 0.399 | 0.398 |
| | | | QB–flat | 0.276 | 0.342 | 0.373 | 0.389 | 0.395 | 0.399 | 0.399 | 0.398 |
| | | | QB–RP | 0.278 | 0.345 | 0.375 | 0.390 | 0.396 | 0.399 | 0.399 | 0.399 |
| | | Blkdiag | GMM | 0.276 | 0.342 | 0.373 | 0.389 | 0.395 | 0.399 | 0.399 | 0.398 |
| | | | QB–flat | 0.276 | 0.342 | 0.373 | 0.389 | 0.395 | 0.399 | 0.399 | 0.398 |
| | | | QB–RP | 0.277 | 0.344 | 0.374 | 0.389 | 0.395 | 0.399 | 0.399 | 0.398 |
| | | HAR | GMM | 0.260 | 0.320 | 0.353 | 0.368 | 0.368 | 0.372 | 0.371 | 0.371 |
| | | | QB–flat | 0.260 | 0.320 | 0.353 | 0.368 | 0.368 | 0.372 | 0.371 | 0.371 |
| | | | QB–RP | 0.264 | 0.325 | 0.358 | 0.374 | 0.373 | 0.377 | 0.377 | 0.377 |
| | LD | Plain | GMM | 0.291 | 0.374 | 0.420 | 0.446 | 0.463 | 0.473 | 0.479 | 0.483 |
| | | | QB–flat | 0.291 | 0.374 | 0.420 | 0.446 | 0.463 | 0.473 | 0.479 | 0.483 |
| | | | QB–RP | 0.294 | 0.379 | 0.426 | 0.452 | 0.470 | 0.480 | 0.486 | 0.490 |
| | | Blkdiag | GMM | 0.291 | 0.374 | 0.420 | 0.446 | 0.463 | 0.473 | 0.479 | 0.483 |
| | | | QB–flat | 0.291 | 0.374 | 0.420 | 0.446 | 0.463 | 0.473 | 0.479 | 0.483 |
| | | | QB–RP | 0.292 | 0.375 | 0.421 | 0.446 | 0.463 | 0.473 | 0.479 | 0.484 |
| | | HAR | GMM | 0.273 | 0.353 | 0.394 | 0.414 | 0.425 | 0.435 | 0.439 | 0.442 |
| | | | QB–flat | 0.273 | 0.353 | 0.394 | 0.414 | 0.425 | 0.435 | 0.439 | 0.442 |
| | | | QB–RP | 0.279 | 0.361 | 0.405 | 0.428 | 0.440 | 0.450 | 0.453 | 0.456 |
| 500 | Level | Plain | GMM | 0.175 | 0.217 | 0.236 | 0.246 | 0.251 | 0.253 | 0.255 | 0.255 |
| | | | QB–flat | 0.175 | 0.217 | 0.236 | 0.246 | 0.251 | 0.253 | 0.255 | 0.255 |
| | | | QB–RP | 0.176 | 0.218 | 0.237 | 0.246 | 0.251 | 0.253 | 0.255 | 0.255 |
| | | Blkdiag | GMM | 0.175 | 0.217 | 0.236 | 0.246 | 0.251 | 0.253 | 0.255 | 0.255 |
| | | | QB–flat | 0.175 | 0.217 | 0.236 | 0.246 | 0.251 | 0.253 | 0.255 | 0.255 |
| | | | QB–RP | 0.175 | 0.217 | 0.236 | 0.246 | 0.251 | 0.253 | 0.255 | 0.255 |
| | | HAR | GMM | 0.168 | 0.210 | 0.228 | 0.236 | 0.240 | 0.242 | 0.242 | 0.241 |
| | | | QB–flat | 0.168 | 0.210 | 0.228 | 0.236 | 0.240 | 0.242 | 0.242 | 0.241 |
| | | | QB–RP | 0.169 | 0.212 | 0.230 | 0.238 | 0.242 | 0.244 | 0.244 | 0.243 |
| | LD | Plain | GMM | 0.183 | 0.238 | 0.266 | 0.283 | 0.293 | 0.299 | 0.304 | 0.307 |
| | | | QB–flat | 0.183 | 0.238 | 0.266 | 0.283 | 0.293 | 0.299 | 0.304 | 0.307 |
| | | | QB–RP | 0.185 | 0.240 | 0.268 | 0.284 | 0.294 | 0.300 | 0.305 | 0.308 |
| | | Blkdiag | GMM | 0.183 | 0.238 | 0.266 | 0.283 | 0.293 | 0.299 | 0.304 | 0.307 |
| | | | QB–flat | 0.183 | 0.238 | 0.266 | 0.283 | 0.293 | 0.299 | 0.304 | 0.307 |
| | | | QB–RP | 0.184 | 0.238 | 0.266 | 0.283 | 0.293 | 0.299 | 0.304 | 0.307 |
| | | HAR | GMM | 0.176 | 0.230 | 0.255 | 0.271 | 0.280 | 0.286 | 0.289 | 0.292 |
| | | | QB–flat | 0.176 | 0.230 | 0.255 | 0.271 | 0.280 | 0.286 | 0.289 | 0.292 |
| | | | QB–RP | 0.178 | 0.233 | 0.258 | 0.274 | 0.283 | 0.289 | 0.293 | 0.295 |
| 1000 | Level | Plain | GMM | 0.123 | 0.153 | 0.167 | 0.174 | 0.178 | 0.180 | 0.180 | 0.181 |
| | | | QB–flat | 0.123 | 0.153 | 0.167 | 0.174 | 0.178 | 0.180 | 0.180 | 0.181 |
| | | | QB–RP | 0.124 | 0.154 | 0.168 | 0.174 | 0.178 | 0.180 | 0.180 | 0.181 |
| | | Blkdiag | GMM | 0.123 | 0.153 | 0.167 | 0.174 | 0.178 | 0.180 | 0.180 | 0.181 |
| | | | QB–flat | 0.123 | 0.153 | 0.167 | 0.174 | 0.178 | 0.180 | 0.180 | 0.181 |
| | | | QB–RP | 0.123 | 0.154 | 0.168 | 0.174 | 0.178 | 0.180 | 0.180 | 0.181 |
| | | HAR | GMM | 0.120 | 0.150 | 0.164 | 0.170 | 0.173 | 0.175 | 0.175 | 0.175 |
| | | | QB–flat | 0.120 | 0.150 | 0.164 | 0.170 | 0.173 | 0.175 | 0.175 | 0.175 |
| | | | QB–RP | 0.120 | 0.151 | 0.166 | 0.171 | 0.174 | 0.176 | 0.176 | 0.176 |
| | LD | Plain | GMM | 0.130 | 0.168 | 0.188 | 0.200 | 0.207 | 0.212 | 0.215 | 0.216 |
| | | | QB–flat | 0.130 | 0.168 | 0.188 | 0.200 | 0.207 | 0.212 | 0.215 | 0.216 |
| | | | QB–RP | 0.131 | 0.168 | 0.189 | 0.201 | 0.208 | 0.212 | 0.215 | 0.216 |
| | | Blkdiag | GMM | 0.130 | 0.168 | 0.188 | 0.200 | 0.207 | 0.212 | 0.215 | 0.216 |
| | | | QB–flat | 0.130 | 0.168 | 0.188 | 0.200 | 0.207 | 0.212 | 0.215 | 0.216 |
| | | | QB–RP | 0.130 | 0.168 | 0.188 | 0.200 | 0.207 | 0.212 | 0.215 | 0.216 |
| | | HAR | GMM | 0.126 | 0.164 | 0.185 | 0.195 | 0.201 | 0.205 | 0.206 | 0.207 |
| | | | QB–flat | 0.126 | 0.164 | 0.185 | 0.195 | 0.201 | 0.205 | 0.206 | 0.207 |
| | | | QB–RP | 0.127 | 0.165 | 0.186 | 0.196 | 0.202 | 0.205 | 0.206 | 0.207 |

Table A.4: Simulation result (4): Pointwise 90% coverage

| $T$ | Spec. | Cov. | Method | Pointwise 90% coverage | | | | | | | |
|---|---|---|---|---|---|---|---|---|---|---|---|
| | | | | $h$ | | | | | | | |
| | | | | 0 | 1 | 2 | 3 | 4 | 5 | 6 | 7 |
| 200 | Level | Plain | GMM | 0.873 | 0.876 | 0.897 | 0.906 | 0.889 | 0.889 | 0.897 | 0.889 |
| | | | QB–flat | 0.874 | 0.877 | 0.897 | 0.903 | 0.888 | 0.889 | 0.898 | 0.890 |
| | | | QB–RP | 0.897 | 0.911 | 0.938 | 0.951 | 0.943 | 0.963 | 0.950 | 0.955 |
| | | Blkdiag | GMM | 0.873 | 0.876 | 0.897 | 0.906 | 0.889 | 0.889 | 0.897 | 0.889 |
| | | | QB–flat | 0.873 | 0.878 | 0.899 | 0.905 | 0.890 | 0.889 | 0.896 | 0.890 |
| | | | QB–RP | 0.900 | 0.921 | 0.934 | 0.942 | 0.943 | 0.945 | 0.940 | 0.942 |
| | | HAR | GMM | 0.837 | 0.847 | 0.854 | 0.864 | 0.848 | 0.853 | 0.852 | 0.847 |
| | | | QB–flat | 0.837 | 0.847 | 0.857 | 0.863 | 0.848 | 0.853 | 0.852 | 0.847 |
| | | | QB–RP | 0.812 | 0.808 | 0.833 | 0.847 | 0.854 | 0.872 | 0.852 | 0.851 |
| | LD | Plain | GMM | 0.897 | 0.877 | 0.882 | 0.883 | 0.878 | 0.887 | 0.903 | 0.902 |
| | | | QB–flat | 0.896 | 0.876 | 0.881 | 0.885 | 0.877 | 0.888 | 0.902 | 0.902 |
| | | | QB–RP | 0.908 | 0.916 | 0.935 | 0.944 | 0.961 | 0.967 | 0.975 | 0.961 |
| | | Blkdiag | GMM | 0.897 | 0.877 | 0.882 | 0.883 | 0.878 | 0.887 | 0.903 | 0.902 |
| | | | QB–flat | 0.897 | 0.875 | 0.881 | 0.883 | 0.878 | 0.887 | 0.901 | 0.903 |
| | | | QB–RP | 0.912 | 0.906 | 0.926 | 0.932 | 0.935 | 0.943 | 0.951 | 0.947 |
| | | HAR | GMM | 0.865 | 0.842 | 0.840 | 0.849 | 0.831 | 0.858 | 0.850 | 0.850 |
| | | | QB–flat | 0.866 | 0.842 | 0.839 | 0.848 | 0.830 | 0.857 | 0.849 | 0.850 |
| | | | QB–RP | 0.828 | 0.822 | 0.826 | 0.849 | 0.872 | 0.876 | 0.872 | 0.851 |
| 500 | Level | Plain | GMM | 0.900 | 0.889 | 0.893 | 0.895 | 0.883 | 0.899 | 0.898 | 0.912 |
| | | | QB–flat | 0.900 | 0.889 | 0.893 | 0.894 | 0.882 | 0.899 | 0.900 | 0.912 |
| | | | QB–RP | 0.883 | 0.876 | 0.891 | 0.914 | 0.924 | 0.935 | 0.947 | 0.946 |
| | | Blkdiag | GMM | 0.900 | 0.889 | 0.893 | 0.895 | 0.883 | 0.899 | 0.898 | 0.912 |
| | | | QB–flat | 0.900 | 0.889 | 0.893 | 0.895 | 0.884 | 0.898 | 0.897 | 0.913 |
| | | | QB–RP | 0.909 | 0.904 | 0.921 | 0.921 | 0.920 | 0.928 | 0.937 | 0.936 |
| | | HAR | GMM | 0.868 | 0.871 | 0.876 | 0.872 | 0.867 | 0.879 | 0.873 | 0.885 |
| | | | QB–flat | 0.868 | 0.872 | 0.874 | 0.873 | 0.867 | 0.879 | 0.874 | 0.885 |
| | | | QB–RP | 0.840 | 0.801 | 0.822 | 0.834 | 0.848 | 0.846 | 0.865 | 0.875 |
| | LD | Plain | GMM | 0.901 | 0.890 | 0.888 | 0.895 | 0.897 | 0.905 | 0.900 | 0.910 |
| | | | QB–flat | 0.901 | 0.890 | 0.888 | 0.895 | 0.898 | 0.906 | 0.898 | 0.910 |
| | | | QB–RP | 0.887 | 0.864 | 0.892 | 0.921 | 0.932 | 0.958 | 0.954 | 0.961 |
| | | Blkdiag | GMM | 0.901 | 0.890 | 0.888 | 0.895 | 0.897 | 0.905 | 0.900 | 0.910 |
| | | | QB–flat | 0.901 | 0.891 | 0.888 | 0.895 | 0.898 | 0.904 | 0.898 | 0.910 |
| | | | QB–RP | 0.911 | 0.904 | 0.910 | 0.918 | 0.930 | 0.941 | 0.943 | 0.941 |
| | | HAR | GMM | 0.878 | 0.863 | 0.861 | 0.866 | 0.859 | 0.869 | 0.870 | 0.893 |
| | | | QB–flat | 0.876 | 0.863 | 0.863 | 0.866 | 0.859 | 0.869 | 0.869 | 0.893 |
| | | | QB–RP | 0.813 | 0.791 | 0.817 | 0.858 | 0.869 | 0.879 | 0.878 | 0.868 |
| 1000 | Level | Plain | GMM | 0.896 | 0.876 | 0.899 | 0.906 | 0.917 | 0.908 | 0.904 | 0.901 |
| | | | QB–flat | 0.897 | 0.874 | 0.899 | 0.906 | 0.917 | 0.908 | 0.903 | 0.902 |
| | | | QB–RP | 0.882 | 0.844 | 0.846 | 0.895 | 0.920 | 0.925 | 0.928 | 0.923 |
| | | Blkdiag | GMM | 0.896 | 0.876 | 0.899 | 0.906 | 0.917 | 0.908 | 0.904 | 0.901 |
| | | | QB–flat | 0.897 | 0.877 | 0.898 | 0.904 | 0.917 | 0.910 | 0.904 | 0.901 |
| | | | QB–RP | 0.909 | 0.880 | 0.905 | 0.923 | 0.930 | 0.933 | 0.925 | 0.919 |
| | | HAR | GMM | 0.889 | 0.863 | 0.880 | 0.883 | 0.904 | 0.894 | 0.889 | 0.887 |
| | | | QB–flat | 0.891 | 0.863 | 0.880 | 0.884 | 0.905 | 0.892 | 0.889 | 0.886 |
| | | | QB–RP | 0.815 | 0.800 | 0.792 | 0.842 | 0.853 | 0.866 | 0.858 | 0.855 |
| | LD | Plain | GMM | 0.890 | 0.899 | 0.910 | 0.898 | 0.909 | 0.905 | 0.897 | 0.918 |
| | | | QB–flat | 0.890 | 0.899 | 0.912 | 0.899 | 0.910 | 0.902 | 0.898 | 0.919 |
| | | | QB–RP | 0.862 | 0.842 | 0.852 | 0.884 | 0.910 | 0.929 | 0.942 | 0.946 |
| | | Blkdiag | GMM | 0.890 | 0.899 | 0.910 | 0.898 | 0.909 | 0.905 | 0.897 | 0.918 |
| | | | QB–flat | 0.889 | 0.898 | 0.912 | 0.899 | 0.909 | 0.905 | 0.896 | 0.918 |
| | | | QB–RP | 0.892 | 0.891 | 0.912 | 0.912 | 0.927 | 0.932 | 0.928 | 0.941 |
| | | HAR | GMM | 0.873 | 0.879 | 0.901 | 0.881 | 0.896 | 0.883 | 0.879 | 0.892 |
| | | | QB–flat | 0.873 | 0.879 | 0.901 | 0.882 | 0.896 | 0.884 | 0.880 | 0.893 |
| | | | QB–RP | 0.843 | 0.805 | 0.823 | 0.848 | 0.862 | 0.889 | 0.899 | 0.904 |

Table A.5: Simulation result (5): Simultaneous 90% coverage

| Spec. | Cov. | Method | $T$ 200 | 500 | 1000 |
|---|---|---|---|---|---|
| Level | Plain | GMM | 0.880 | 0.867 | 0.903 |
| | | QB–flat | 0.876 | 0.867 | 0.904 |
| | | QB–RP | 0.949 | 0.931 | 0.888 |
| | Blkdiag | GMM | 0.894 | 0.898 | 0.916 |
| | | QB–flat | 0.871 | 0.868 | 0.899 |
| | | QB–RP | 0.938 | 0.937 | 0.928 |
| | HAR | GMM | 0.755 | 0.809 | 0.845 |
| | | QB–flat | 0.750 | 0.800 | 0.835 |
| | | QB–RP | 0.754 | 0.767 | 0.775 |
| LD | Plain | GMM | 0.891 | 0.891 | 0.899 |
| | | QB–flat | 0.883 | 0.895 | 0.902 |
| | | QB–RP | 0.963 | 0.939 | 0.904 |
| | Blkdiag | GMM | 0.914 | 0.921 | 0.925 |
| | | QB–flat | 0.884 | 0.895 | 0.902 |
| | | QB–RP | 0.942 | 0.937 | 0.927 |
| | HAR | GMM | 0.772 | 0.824 | 0.857 |
| | | QB–flat | 0.774 | 0.820 | 0.860 |
| | | QB–RP | 0.770 | 0.794 | 0.835 |

# Appendix B: Instrument Diagnostics

## B.1 First-stage relevance diagnostics

**First-stage regressions** To assess first-stage relevance, we estimate the auxiliary regressions

$$\begin{aligned} r_t^{wind} &= \xi_{1,1} q_t^{wind} + \xi_{1,2} q_t^{solar} + \boldsymbol{\lambda}_1^\top \check{\boldsymbol{x}}_t + v_{1,t}, \\ r_t^{solar} &= \xi_{2,1} q_t^{wind} + \xi_{2,2} q_t^{solar} + \boldsymbol{\lambda}_2^\top \check{\boldsymbol{x}}_t + v_{2,t}, \end{aligned}$$

where $\check{\boldsymbol{x}}_t$ contains all included controls in the baseline LP–IV specification:

$$\begin{aligned} \check{\boldsymbol{x}}_t &= \left(1, \Delta y_{t-1}, ..., \Delta y_{t-L}, \sin(2\pi 1 s_t), ..., \sin(2\pi N s_t), \right. \\ & \quad \left. \cos(2\pi 1 s_t), ..., \cos(2\pi N s_t), \check{\boldsymbol{x}}_t^\top \right)^\top . \end{aligned}$$

Equivalently,

$$\boldsymbol{r}_t = \boldsymbol{\Xi} \boldsymbol{q}_t + \boldsymbol{\Lambda} \check{\boldsymbol{x}}_t + \boldsymbol{v}_t,$$

where

$$\boldsymbol{q}_t = \begin{pmatrix} q_t^{wind} \\ q_t^{solar} \end{pmatrix}, \quad \boldsymbol{\Xi} = \begin{pmatrix} \xi_{1,1} & \xi_{1,2} \\ \xi_{2,1} & \xi_{2,2} \end{pmatrix},$$

$$\boldsymbol{\Lambda} = \begin{pmatrix} \boldsymbol{\lambda}_1^\top \\ \boldsymbol{\lambda}_2^\top \end{pmatrix} \quad \boldsymbol{v}_t = \begin{pmatrix} v_{1,t} \\ v_{2,t} \end{pmatrix}.$$

The diagonal elements of $\boldsymbol{\Xi}$ measure own first-stage relevance, while the off-diagonal elements diagnose cross-loading between weather-potential instruments and renewable-generation variables.

**Partial $R^2$ for excluded instruments** The partial $R^2$ measures the incremental explanatory power of the excluded renewable-potential instruments, conditional on the included controls, separately for the wind and solar first-stage equations. For each endogenous variable $r_t^j$, $j \in \{wind, solar\}$, define

$$\boldsymbol{M}_{\check{\boldsymbol{X}}} = \boldsymbol{I}_T - \check{\boldsymbol{X}} \left( \check{\boldsymbol{X}}^\top \check{\boldsymbol{X}} \right)^{-1} \check{\boldsymbol{X}}^\top, \quad \check{\boldsymbol{X}} = (\check{\boldsymbol{x}}_1, ..., \check{\boldsymbol{x}}_T)^\top .$$

Let

$$\boldsymbol{r}^j = \left( r_1^j, ..., r_T^j \right)^\top, \quad \boldsymbol{Q} = (\boldsymbol{q}_1, ..., \boldsymbol{q}_T)^\top .$$

The partial $R^2$ of the excluded instruments for endogenous variable $j$ is

$$\tilde{R}_j^2 = 1 - \frac{\boldsymbol{r}^{j\top} \boldsymbol{M}_{\boldsymbol{Q}, \check{\boldsymbol{X}}} \boldsymbol{r}^j}{\boldsymbol{r}^{j\top} \boldsymbol{M}_{\check{\boldsymbol{X}}} \boldsymbol{r}^j},$$

where

$$\boldsymbol{M}_{\boldsymbol{Q}, \check{\boldsymbol{X}}} = \boldsymbol{I}_T - \left[ \begin{array}{cc} \boldsymbol{Q} & \check{\boldsymbol{X}} \end{array} \right] \left( \left[ \begin{array}{cc} \boldsymbol{Q} & \check{\boldsymbol{X}} \end{array} \right]^\top \left[ \begin{array}{cc} \boldsymbol{Q} & \check{\boldsymbol{X}} \end{array} \right] \right)^{-1} \left[ \begin{array}{cc} \boldsymbol{Q} & \check{\boldsymbol{X}} \end{array} \right]^\top .$$

**Excluded instrument Wald $F$-statistic** The excluded-instrument Wald/$F$-type statistic assesses whether the renewable-potential instruments predict each endogenous variable after conditioning on the included controls. With multiple endogenous variables, equation-by-equation first-stage statistics are informative but not sufficient, because weak identification may also arise from near rank reduction in the first-stage coefficient matrix. For each first-stage equation, we test

$$\begin{aligned} H_0^{wind} &: \ \xi_{1,1} = \xi_{1,2} = 0, \\ H_0^{solar} &: \ \xi_{2,1} = \xi_{2,2} = 0. \end{aligned}$$

Let

$$\hat{\boldsymbol{\xi}}_j = \left( \hat{\xi}_{j,1}, \hat{\xi}_{j,2} \right)^\top,$$

where $j = 1$ corresponds to wind generation and $j = 2$ corresponds to solar generation. Let $\widehat{\text{Var}}\left(\hat{\boldsymbol{\xi}}_j\right)$ denote the estimated covariance matrix of the two excluded-instrument coefficients in the $j$th first-stage regression. The HAR-robust Wald/$F$-type relevance statistic is

$$F_j = \frac{1}{2} \hat{\boldsymbol{\xi}}_j^\top \widehat{\text{Var}}\left(\hat{\boldsymbol{\xi}}_j\right)^{-1} \hat{\boldsymbol{\xi}}_j .$$

Table B.1: First-stage relevance check

| | | | Regressand | |
|---|---|---|---|---|
| | | | $r_t^{wind}$ | $r_t^{solar}$ |
| DK1 | Coefficient estimate | $WindPot_t$ | 0.336 | -0.016 |
| | | | (0.035) | (0.030) |
| | | $SolarPot_t$ | 0.073 | 0.913 |
| | | | (0.029) | (0.068) |
| | Partial-$R^2$ | | 0.100 | 0.301 |
| | Wald/$F$-type relevance statistic | | 45.578 | 108.468 |
| | Min. singular value of $\hat{\Xi}$ | | | 0.336 |
| DK2 | Coefficient estimate | $WindPot_t$ | 0.633 | -0.035 |
| | | | (0.031) | (0.033) |
| | | $SolarPot_t$ | 0.100 | 0.915 |
| | | | (0.032) | (0.069) |
| | Partial-$R^2$ | | 0.246 | 0.269 |
| | Wald/$F$-type relevance statistic | | 210.524 | 94.900 |
| | Min. singular value of $\hat{\Xi}$ | | | 0.633 |

Note: A number in parentheses is a HAR standard error.

**Multivariate first-stage relevance diagnostic** Because there are two endogenous variables, equation-by-equation first-stage diagnostics may fail to detect near rank deficiency in the first-stage coefficient matrix. We therefore report the smallest singular value of the standardized excluded-instrument coefficient matrix as a descriptive diagnostic for joint relevance.[1] We do not interpret this quantity as a formal weak-instrument test statistic. Because renewable generation and renewable-potential instruments are standardized before estimation, the singular values are reported on a standardized scale.

Table B.1 reports the first-stage relevance diagnostics. In both bidding zones, the excluded renewable-potential instruments strongly predict the corresponding renewable-generation variables after conditioning on the baseline controls. The partial $R^2$ values range from 0.100 to 0.301, and the HAR-robust Wald/$F$-type statistics are large. The smallest singular values of the standardized first-stage coefficient matrix are also clearly separated from zero. Together, these diagnostics indicate that the excluded instruments provide substantial identifying variation for both wind and solar generation.

## B.2 Placebo diagnostics

The instruments should not be strongly associated with predetermined market conditions after conditioning on controls. To examine this requirement, we regress predetermined variables on wind and solar potential using the same baseline controls. This diagnostic is especially relevant for solar because sunny conditions may be correlated with residual demand conditions. If solar potential predicts lagged prices or load, the estimated solar response may partly reflect residual seasonality or demand variation rather than a clean renewable supply shock.

We examine whether the excluded meteorological instruments predict predetermined electricity-market conditions,

$$b_t \in \left\{y_{t-1}, \Delta y_{t-1}, Load_{t-1}, \Delta Load_{t-1}, r_{t-1}^{wind}, r_{t-1}^{solar}\right\},$$

where $Load_t$ denotes total electricity load. For each predetermined variable $b_t$, we estimate

$$b_t = \boldsymbol{\phi}_b^\top \boldsymbol{q}_t + \boldsymbol{\psi}_b^\top \dot{\boldsymbol{x}}_t + v_t,$$

where $\boldsymbol{\phi}_b = (\phi_{b,1}, \phi_{b,2})^\top$ and $\boldsymbol{\psi}_b$ are coefficient vectors. The control vector $\dot{\boldsymbol{x}}_t$ is the same as $\check{\boldsymbol{x}}_t$; except that when $b_t = \Delta y_{t-1}$, $\Delta y_{t-1}$ is removed from $\check{\boldsymbol{x}}_t$.

We report a joint diagnostic for the null hypothesis

$$H_{0,b}: \ \boldsymbol{\phi}_b = \mathbf{0}_2.$$

The robust Wald $p$-value is

$$p_b = 1 - F_{\chi_2^2}\left(\hat{\boldsymbol{\phi}}_b^\top \widehat{\mathrm{Var}}\left(\hat{\boldsymbol{\phi}}_b\right)^{-1} \hat{\boldsymbol{\phi}}_b\right),$$

[1] See Cragg & Donald (1993); Kleibergen & Paap (2006); Sanderson & Windmeijer (2016) for a related discussion.

Table B.2: Placebo check

| | | | Regressand | | | | | |
|---|---|---|---|---|---|---|---|---|
| | | | $y_{t-1}$ | $\Delta y_{t-1}$ | $Load_{t-1}$ | $\Delta Load_{t-1}$ | $r_{t-1}^{wind}$ | $r_{t-1}^{solar}$ |
| DK1 | Coefficient estimate | $WindPot_t$ | -1.501 | -1.468 | -6.639 | 15.509 | -0.467 | 0.012 |
| | | | (0.829) | (0.400) | (6.518) | (6.229) | (1.064) | (0.064) |
| | | $SolarPot_t$ | 0.119 | -0.61 | 18.773 | 2.664 | 1.603 | 0.185 |
| | | | (1.155) | (0.514) | (8.190) | (7.779) | (1.220) | (0.103) |
| | Partial-$R^2$ | | 0.002 | 0.005 | 0.004 | 0.003 | 0.001 | 0.004 |
| | Wald $p$-statistic | | 0.170 | 0.001 | 0.055 | 0.036 | 0.345 | 0.200 |
| DK2 | Coefficient estimate | $WindPot_t$ | -1.946 | -0.807 | -6.532 | 6.712 | 1.121 | 0.011 |
| | | | (0.981) | (0.494) | (4.361) | (3.164) | (0.294) | (0.029) |
| | | $SolarPot_t$ | -0.858 | 0.641 | 9.488 | 0.173 | 0.583 | 0.132 |
| | | | (1.020) | (0.485) | (4.366) | (2.488) | (0.307) | (0.045) |
| | Partial-$R^2$ | | 0.003 | 0.002 | 0.005 | 0.002 | 0.006 | 0.008 |
| | Wald $p$-statistic | | 0.102 | 0.136 | 0.014 | 0.105 | 0.000 | 0.012 |

Note: A number in parentheses is a HAR standard error.

where $\widehat{\text{Var}}\left(\hat{\boldsymbol{\phi}}_b\right)$ denotes the estimated covariance matrix of $\hat{\boldsymbol{\phi}}_b$ and $F_{\chi_2^2}(\cdot)$ is the cumulative distribution function of a chi-squared distribution with two degrees of freedom.

The partial $R^2$ for renewable potential is defined analogously to Section B.1,

$$\tilde{R}_b^2 = 1 - \frac{\boldsymbol{b}^\top \boldsymbol{M}_{\boldsymbol{Q},\dot{\boldsymbol{X}}} \boldsymbol{b}}{\boldsymbol{b}^\top \boldsymbol{M}_{\dot{\boldsymbol{X}}} \boldsymbol{b}},$$

where

$$\boldsymbol{b} = (b_1, ..., b_T)^\top,$$

$$\boldsymbol{M}_{\dot{\boldsymbol{X}}} = \boldsymbol{I}_T - \dot{\boldsymbol{X}}\left(\dot{\boldsymbol{X}}^\top \dot{\boldsymbol{X}}\right)^{-1} \dot{\boldsymbol{X}}^\top,$$

$$\boldsymbol{M}_{\boldsymbol{Q},\dot{\boldsymbol{X}}} = \boldsymbol{I}_T - \left[\begin{array}{cc} \boldsymbol{Q} & \dot{\boldsymbol{X}} \end{array}\right]\left(\left[\begin{array}{cc} \boldsymbol{Q} & \dot{\boldsymbol{X}} \end{array}\right]^\top \left[\begin{array}{cc} \boldsymbol{Q} & \dot{\boldsymbol{X}} \end{array}\right]\right)^{-1} \left[\begin{array}{cc} \boldsymbol{Q} & \dot{\boldsymbol{X}} \end{array}\right]^\top,$$

$$\dot{\boldsymbol{X}} = (\dot{\boldsymbol{x}}_1, ..., \dot{\boldsymbol{x}}_T)^\top.$$

Table B.2 reports the placebo diagnostics. The partial $R^2$ values are uniformly small, indicating that renewable-potential instruments add little explanatory power for predetermined market variables after conditioning on the baseline controls. Some Wald $p$-values are nevertheless small, especially for lagged renewable generation in DK2. This suggests that the instruments are better interpreted as weather-state instruments than as pure weather-innovation instruments, reflecting serial correlation in wind and solar conditions. These results motivate the lag-augmented robustness specification reported in Section 5.6, while the small partial $R^2$ values suggest that the baseline instruments are not strongly associated with pre-existing market conditions.

Figure B.1: Estimated lead-placebo IRF

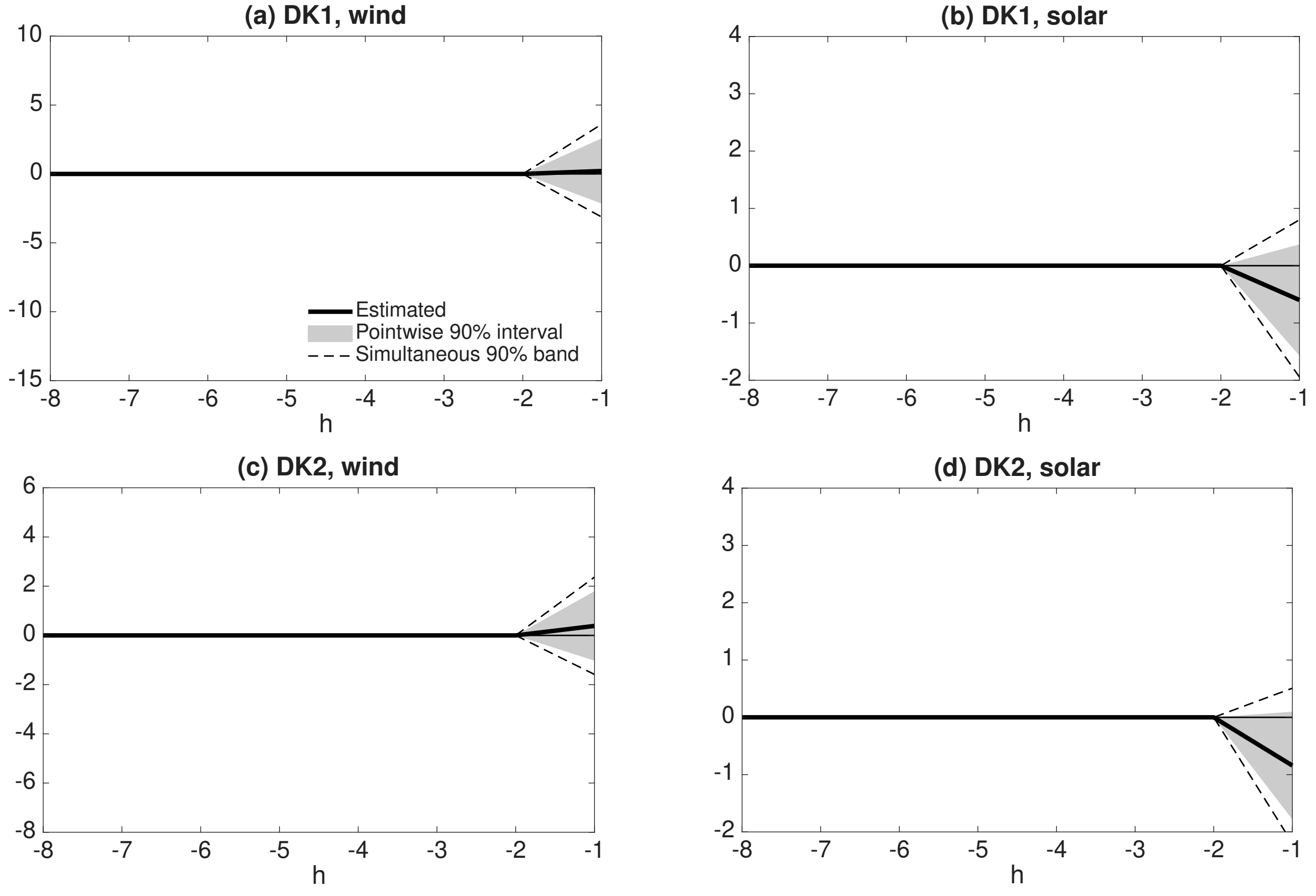

## Appendix C: Robustness Check

Figure C.1: Robustness check (1): Level specification

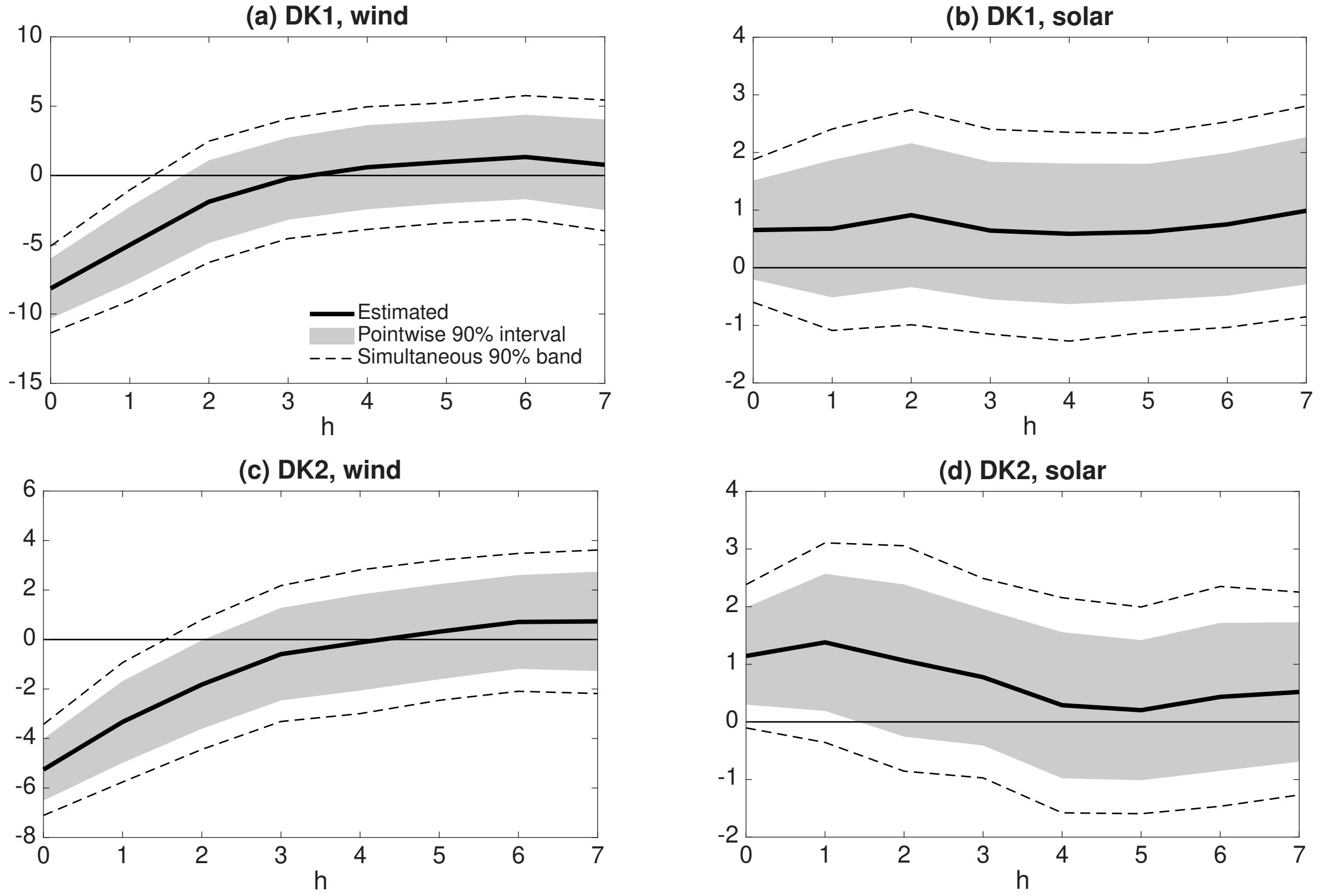

Figure C.2: Robustness check (2): $L = 14$

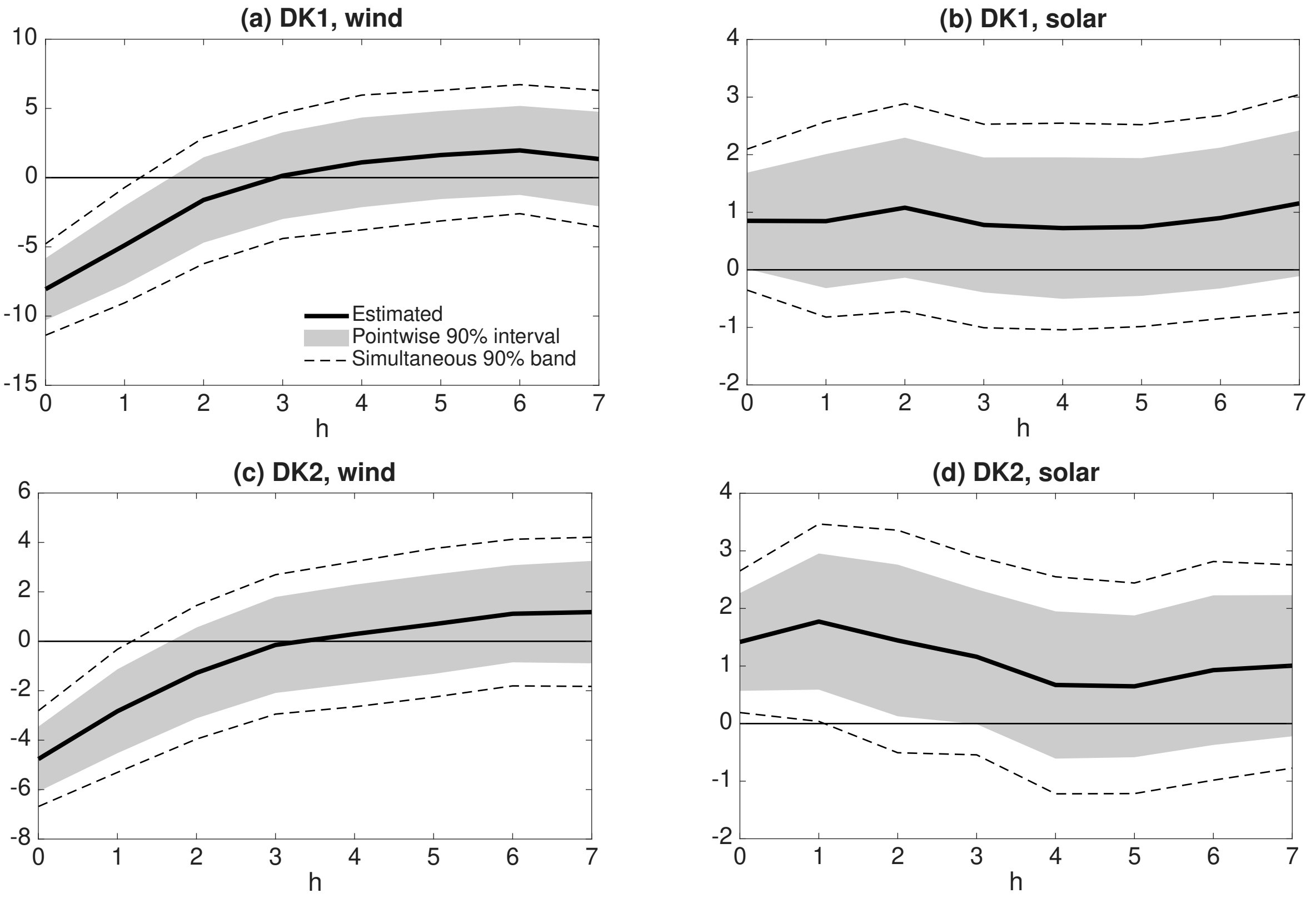

Figure C.3: Robustness check (3): $N = 8$

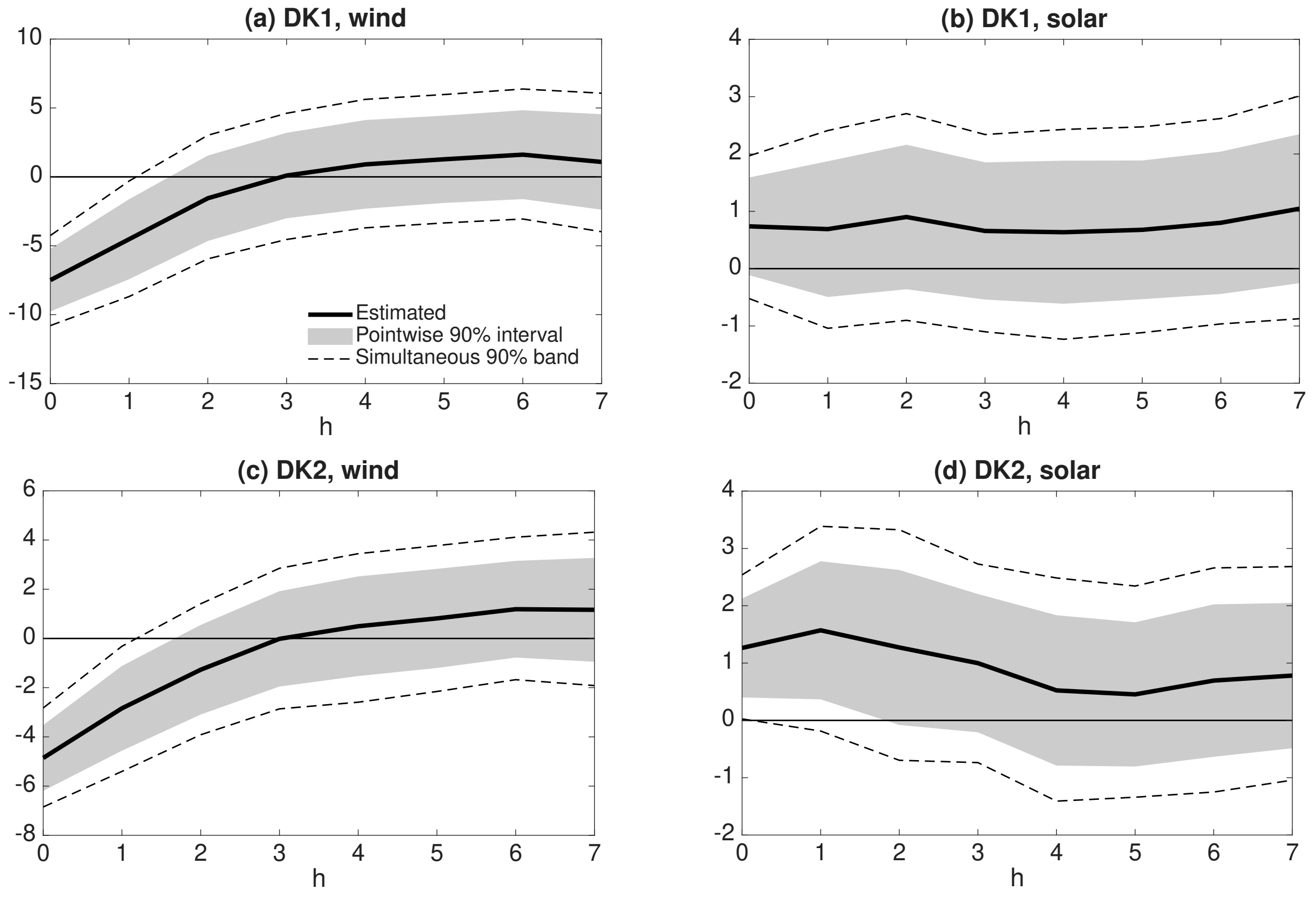

Figure C.4: Robustness check (4): $\rho = 2$

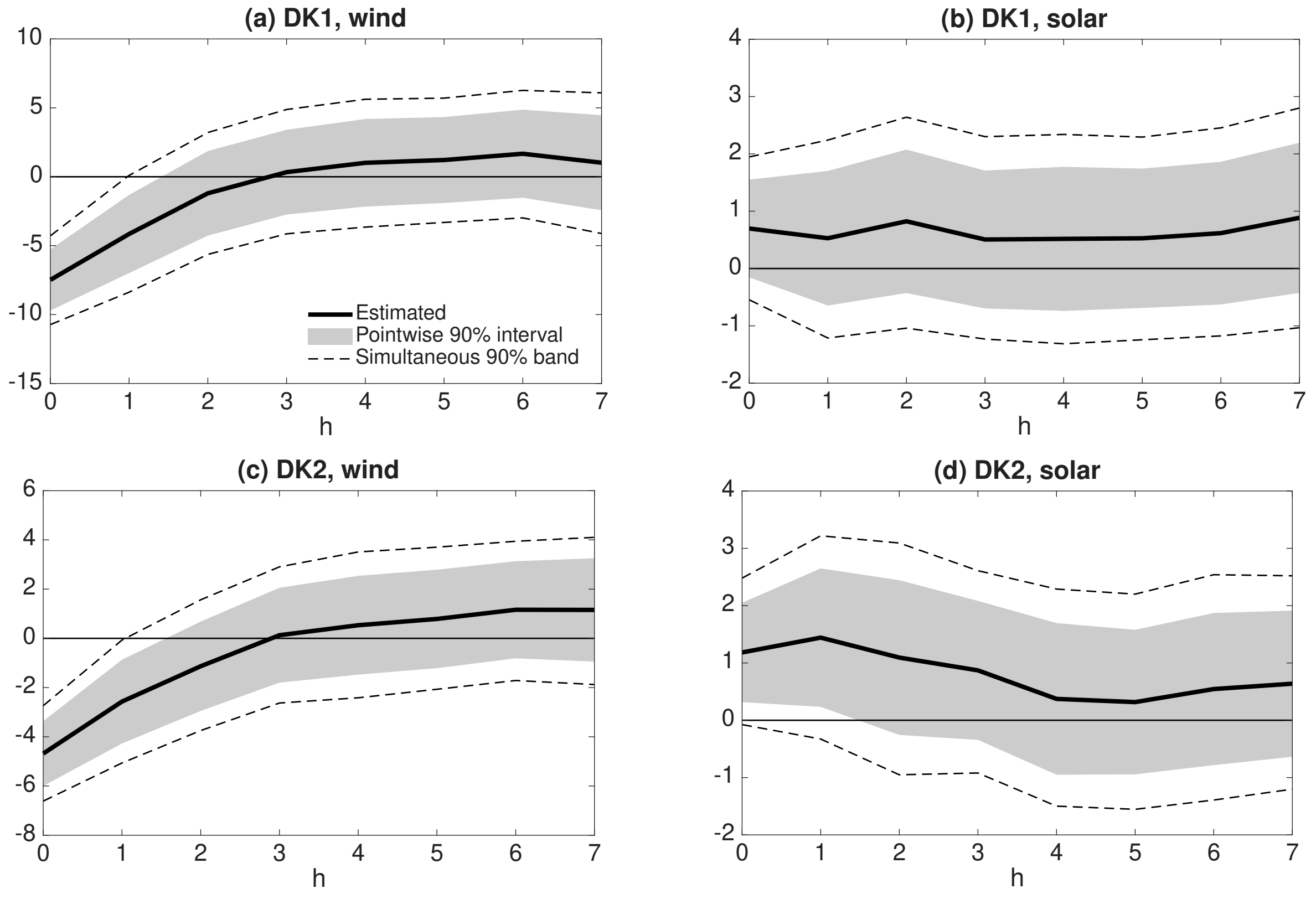

Figure C.5: Robustness check (5): $\rho = 8$

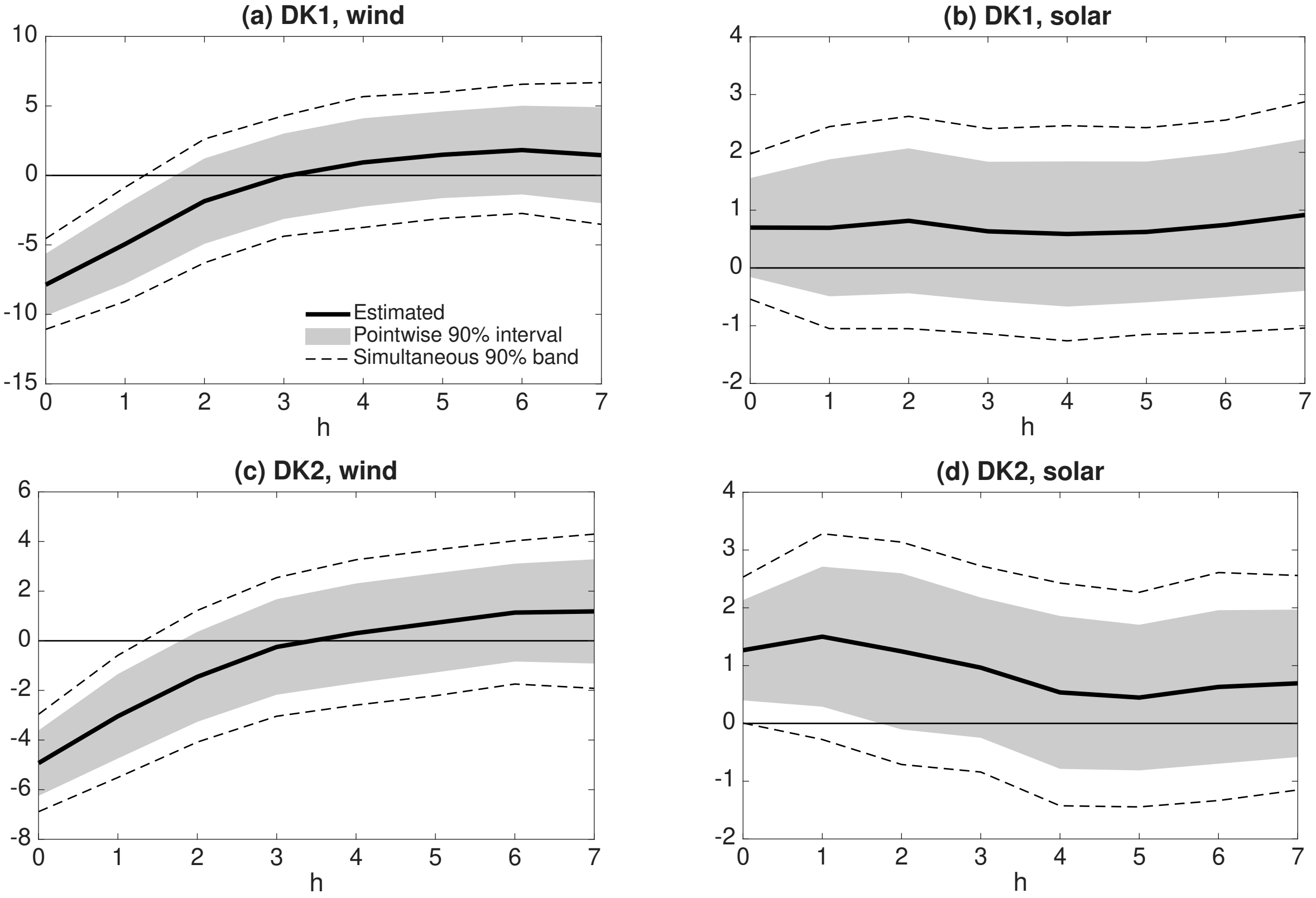

Figure C.6: Robustness check (6): $\kappa = 0.1$

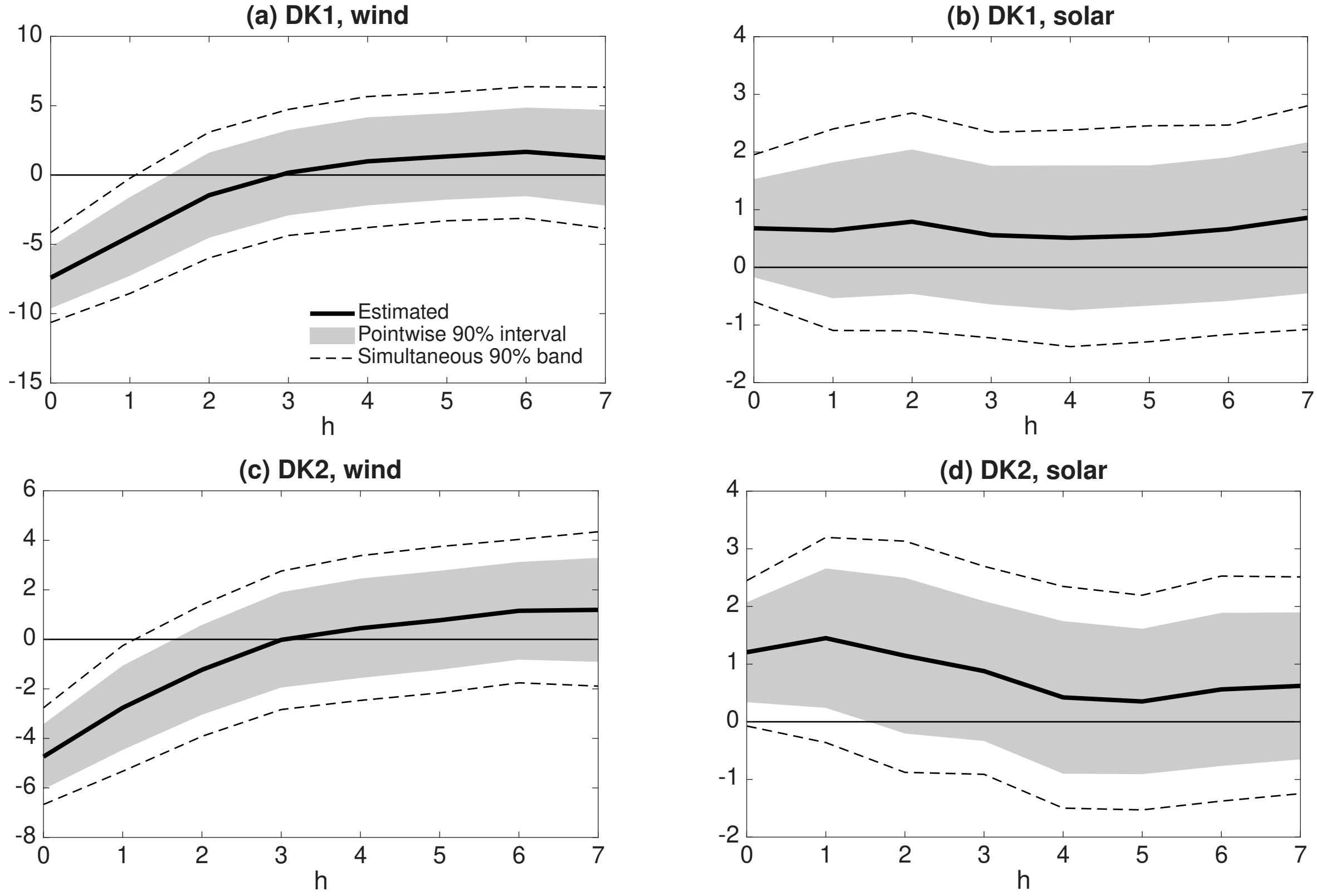

Figure C.7: Robustness check (7): $\kappa = 10$

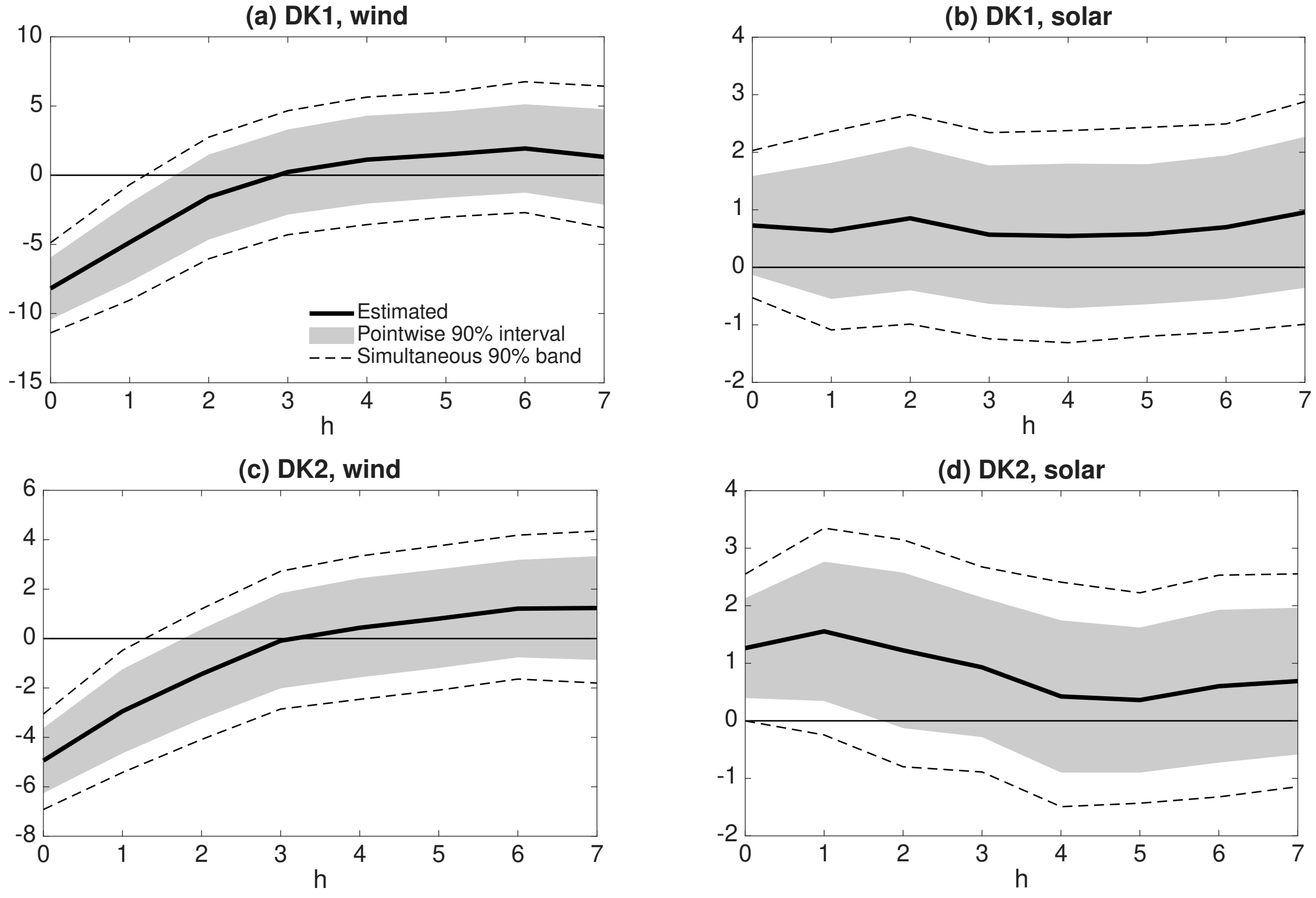

Figure C.8: Robustness check (8): Lag-augmented specification

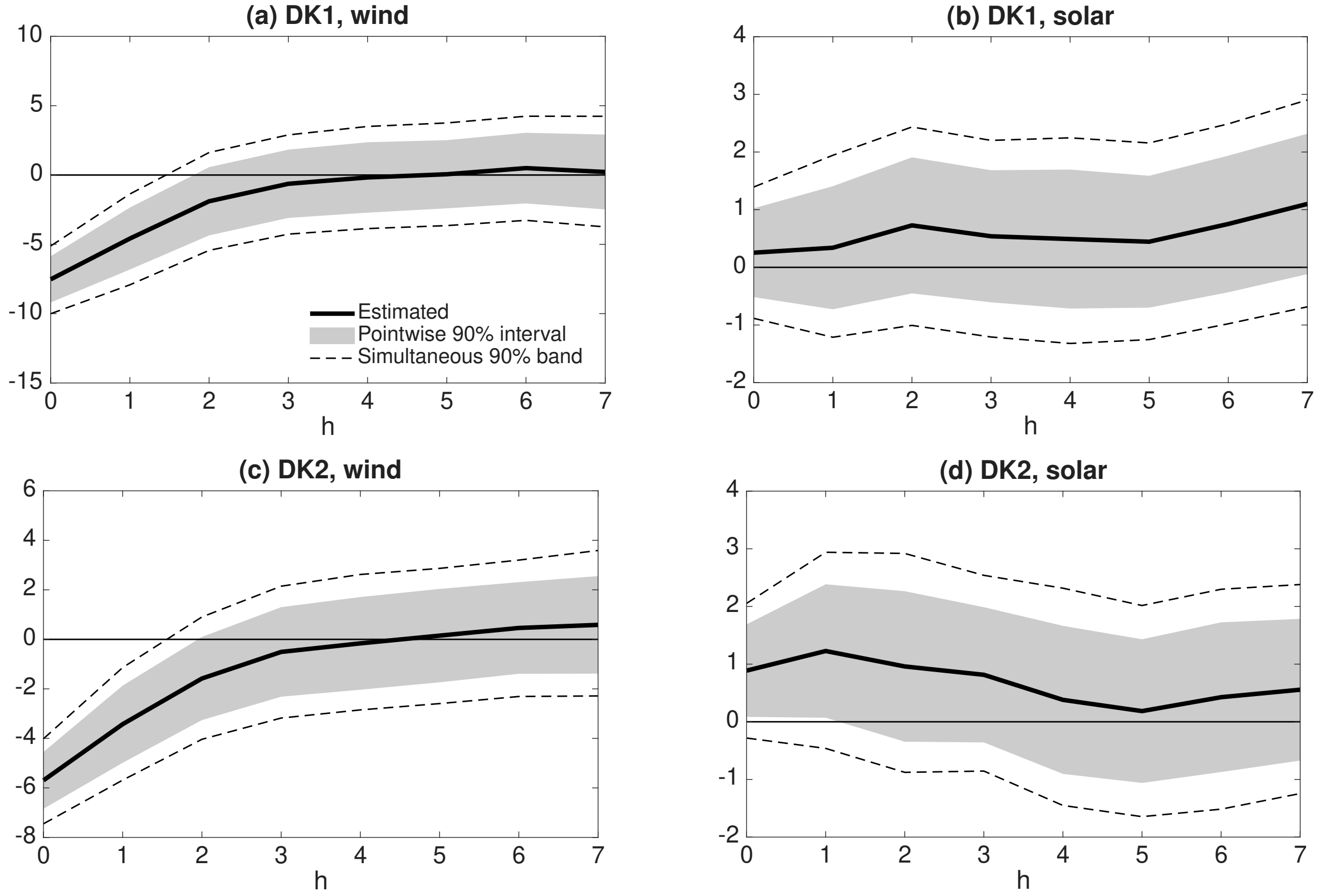

Figure C.9: Robustness check (9): Flat prior

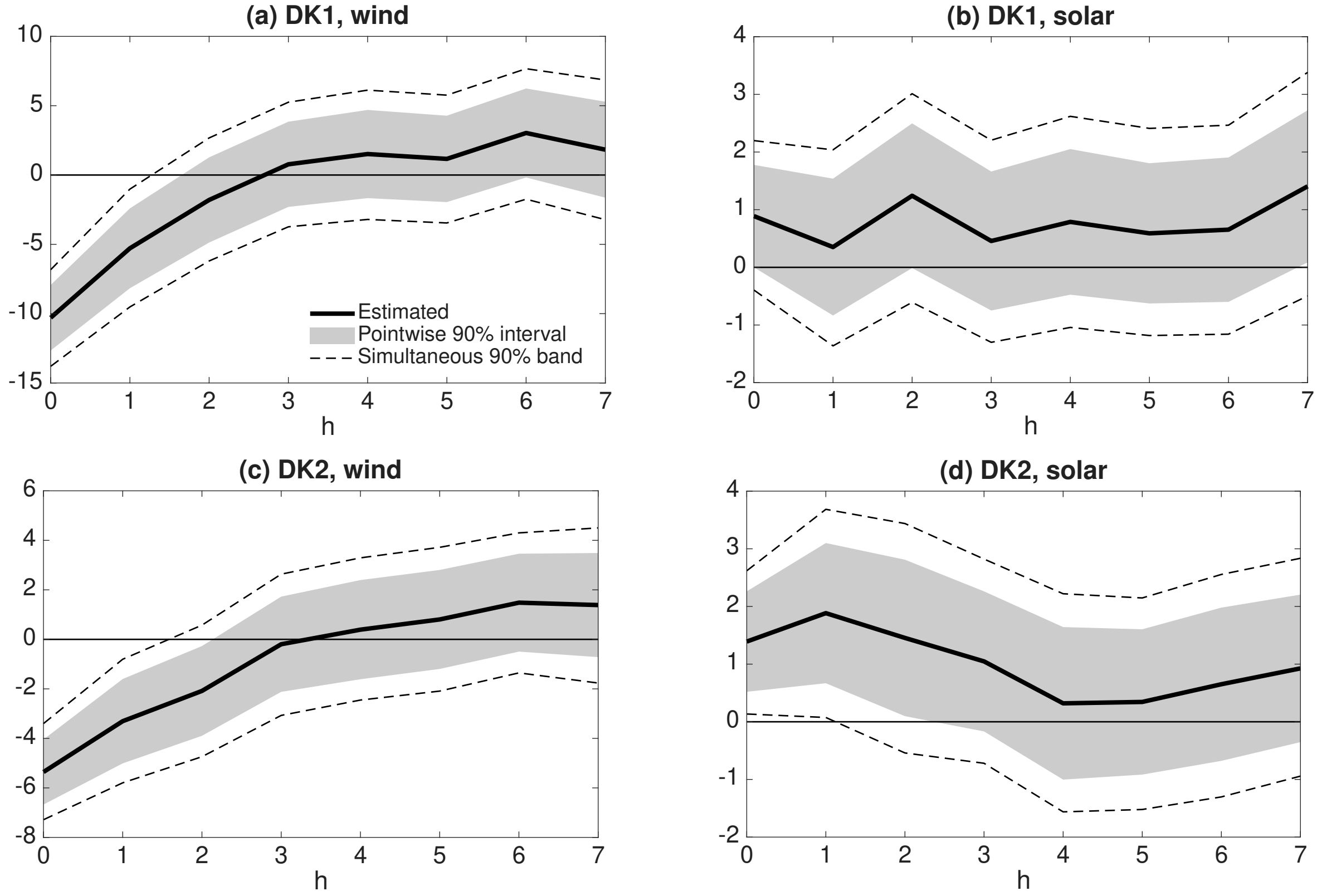


\end{document}